\documentclass[%
aps,
onecolumn,
groupedaddress,
showkeys,
amsmath,
amssymb,
]{revtex4-2}
\usepackage{graphicx}
\usepackage{dcolumn}
\usepackage{bm}
\usepackage{caption}
\usepackage{subfloat}
\usepackage{subfig}
\usepackage{float}

\renewcommand{\theequation}{\arabic{equation}}
\begin{document}

\preprint{APS/123-QED}
\title{Two-Dimensional Materials-Based Josephson Junctions}
\author{Hamidreza Simchi}
\email{simchi@alumni.iust.ac.ir}
\affiliation {Department of Physics, Iran University of Science and Technology, Narmak, Tehran 16844, Iran} 
\maketitle
\date{\today}

\noindent  We consider a two-dimensional monolayer MoS${}_{2}$-based Josephson junction which is composed by an intermediate semiconductor flake and the semi-infinite topological and non-topological superconductor leads and study its quantum transport properties by using the tight-binding non-equilibrium Green function method. By introducing a simple tight-binding model, it is shown that, when the absolute value of chemical potential is much smaller than the superconductor paring potential, the Majorana zero modes, whose Chern number is two, are formed in the topological leads. Also, we show that, in Josephson junction with ordinary superconductor leads, the Josephson current has sinusoidal behavior (due to forming the Andreev bound states (ABS)), when the absolute value of energy of carriers (and the chemical potential) is much smaller (greater) than the superconductor pairing potential. Of course, for Josephson junction with topological superconductor leads, it is shown that the ABS are not formed and in consequence the related Josephson current is zero. Therefore, one can consider the two-dimensional monolayer MoS${}_{2}$-based Josephson junction as a two-state switch which is in open-state (due to ABS) when $\mu \ge 0.8$ eV and is in close-state (due to Majorana) when $\mu <0.8$ or $\mu =0$, i.e., ABS cannot mimic the Majorana state, as zero-bias conductance.

\section{Introduction}
\noindent In every scientific revolution, not only the weaknesses and limitations of previous theories overcome, but also the new concepts and tools are made available to scientists [1]. After presentation of the theory of quantum physics, new concepts specific to this theory, such as the state functions, their superposition and tunneling through a potential barrier, entanglement between the state functions, and the reduction of state functions in the measurement process, became available to scientists. The quantum technology is an emerging technology whose development is based on the direct use of these new concepts.

After introducing the theory of quantum physics, two questions are asked: Can quantum physics be simulated by a computer? And is it possible to have a quantum computer, that is, a computer that uses quantum bits (superposition of two orthogonal quantum state functions which have two opposite eigenvalues and satisfies the Divincenzo's requirements [2]) instead of classical bits and that operates according to the laws of quantum physics [3]?

Among various research activities aimed at providing positive answers to these questions, three research fields called, photonic computers [4], superconducting computers [5], and topological computers [6], have received more attention than other fields in recent years. In designing and building photonic computers, people use the state function and polarization (spin) of the photon [4, 7-12]. Due to the interaction of the system with the surrounding environment, the impurity atoms and the crystal defects and the intrinsic nature of spin of photons, an unwanted type of decoherence phenomenon is created in the system, which is the root of the physical (not computational) faults. For this reason, some peoples try to reduce the amount of interactions and to use extra high pure materials which have zero impurities and crystal defects, approximately [13-15].

In parallel with these activities, another group of researchers, due to the problems caused by the decoherence phenomenon in photonic computers, prefer to choose the superconducting technology and to use the superconducting quantum circuits. One of the most important devices in this technology is the Josephson junction. The nonlinearity of the Josephson junction has enabled people to design and build the superconducting quantum bits whose properties do not depend on the intrinsic spin degrees of freedom [5]. 

But, since producing, preparing, processing of materials, and creating the low temperatures required for superconducting devices and isolating them from the surrounding environment are not only difficult but also expensive, some other research groups prefer to choose the topological quantum computer approach [6]. The cornerstone of this approach is the concept and relationship between symmetry and topology in condensed matter physics. The topological properties that define the different phases of matter are inherently quantum and have a sensitive (fuzzy and subtle) relationship to the concept of entanglement [6]. It is shown that the single-particle Hamiltonian can be classified into ten groups based on the time-reversal, particle-hole, and chiral symmetries (Atland-Zirnbauer table) [6,16,17]. Also, based on the ten groups classification and the spatial dimension, a periodic table for topological insulators and superconductors has been presented [6,16,17].  Also, the $Z_2$ topological invariant, which is called the Majorana number, has been introduced by Kitaev [18,19]. The solid-state Majorana wave function is attributed to electron and its antiparticle (hole), simultaneously.  

The Majorana particle is neither a Boson nor a Fermion, but an Anyon. If we assume that over time, the Anyons in a system exchange their places with each other, and their exchange are represented by the Pauli operator, ${\sigma }_I$, which exchanges the place of particle $I$ with the place of particle $I+1$, it can be shown that by using the properties of the braiding group, the time evolution of these exchanges can be studied [20,21]. In three dimensions, the exchange group is the group,$S_N$ , and in two dimensions, it is the braiding group, $B_N$ [20]. The one-dimensional representation of the braiding group is an Abelian representation and its anyon is called Abelian anyon. Representations of the braiding group in more than one dimension is called non-Abelian representation and its anyon is called non-Abelian anyon [20]. People have shown that how to perform fault-tolerant quantum computations with anyons [18,21]. Of course, the core stone of these calculations is the Majorana zero mode (MZM), whose time evolution in this two-dimensional lattice occurs by the Pauli operators,${\sigma }_x$ , and,${\sigma }_z$ , and based on the rules of braiding group. So from a theoretical perspective, people had been able to solve the problem of fault-tolerant quantum computation in the topological quantum computers approach, but the new problem they had encountered was the experimental observation of the MZM in solid-state physics [18,21].

\noindent        Many attempts have been made to observe the MZM [22-29]. Although there have been some comments regarding the observation of the Majorana mode zero [30], the creation of the Majorana1-chip by IBM [31] and the presentation of a roadmap for the construction of the first topological quantum computer [32] have dispelled doubts in this regard and increased global efforts to achieve fault-tolerant computers. In these experiments, epitaxial InAs/Al nanowire [22,25], Sub-monolayer Fe on single crystal Pb(110) [23],  superconductor Nb bar/quantum anomalous Hall insulator (QAHI) Cr${}_{0.12}$Bi${}_{0.26}$Sb${}_{0.62}$ thin film [24],  Al/InGaAs-InAs/Al structure [26], crystal structure of FeTe${}_{0.55}$Se${}_{0.45}$ [27], Cr/Al contacts with Al-shell wire [28], superconductor Al(Pb)/magnetic insulator EuS [29], Nb wire/QAHI (Bi, Sb)${}_{1.85}$Cr${}_{0.15}$Te${}_{3}$ [30], and superconductor Al wire/quantum dot InAs [31] have been used. Therefore, superconductor/normal metal (SC/NM), superconductor/semiconductor (SC/SM), superconductor/insulator (SC/I) structures and Josphson junction as SC/NM/SC, SC/SM/SC, and SC/I/SC with topological properties of MZM have been considered. Of course, since the trivial Andreev-bound states mimic the Majorana-like signatures as a zero-bias conductance peaks, more robust verification methods, advanced material engineering, and machine-learning-assisted disorder mitigation techniques should be used.

\noindent      Now, the question that can be raised is that given the prediction of topological properties in two-dimensional graphene-like materials [33-39], their superconducting properties [40-43] and the superconductor/two-dimensional material junction [44-47], can a two-dimensional material-based Josephson junction be designed and fabricated for building a switch which is required for manufacturing the topological quantum computers? In this paper, we seek to provide a positive answer to this question.

In this paper, we consider a two-dimensional monolayer MoS${}_{2}$-based Josephson junction and study its quantum transport properties by using the tight-binding non-equilibrium Green function (TB-NEGF) method. By introducing a simple tight-binding model, it is shown that, when the absolute value of chemical potential is much smaller than the superconductor paring potential, the Majorana zero modes, whose Chern number is two, are formed in the topological leads. We show that, in Josephson junction with non-topological superconductor leads, the Josephson current has sinusoidal behavior, due to forming the Andreev bound states (ABS), when the absolute value of carrier energies (and the chemical potential) is much smaller (greater) than the superconductor pairing potential. Of course, for Josephson junction with topological superconductor leads, ABS are not formed and in consequence the related Josephson current is zero. Therefore, one can consider the two-dimensional monolayer MoS${}_{2}$-based Josephson junction as a two-state switch which is in open-state when $\mu \ge 0.8$ eV and is close-state when $\mu <0.8$ or $\mu =0$ (for topological superconductor leads). Therefore, the open-state is attributed to the Majorana state while the close-state is referred to the ABS i.e., ABS cannot mimic the Majorana state, as zero-bias conductance.  

\noindent        Therefore, based on these results and in agreement with the global approach to developing the hybrid two-dimensional material (2DM)-CMOS technology [51], it can be concluded that the use of two-dimensional material-based Josephson junctions may be put on the research roadmap for designing and manufacturing the topological switch which is required for manufacturing the topological computers. It should be noted that, in this paper, for extracting the Hamiltonian of channel and leads, we always consider the  non- Lavazier sublattice of Mo-atoms in the infinite zigzag nanoribbon of MoS${}_{2}$, such that, the $x$-direction is in parallel with the zigzag edge. Also, we consider an effective discrete lattice description of the nanoribbons and the Josephson junction and not an atomic model [51].

\noindent      The structure of article is as follows. Introduction and tight-binding Hamiltonian of semiconductor channel (intermediate layer of Josephson junction) are provided in section I and II, respectively. The Hamiltonian of non-topological (ordinary) and topological superconducting leads are presented in section III and IV. In section V, the Hamiltonian model of Josephson junction is discussed and the results of numerical calculation related to the sections II to V are presented and discussed in section VI. The conclusion is presented in section VII. In Appendix A, the detailed explanations of TB-NEGF method is provided.

\section{Tight-binding model of semiconductor channel}

\noindent 
\noindent         The molybdenum (Mo) atoms in two-dimensional monolayer MoS${}_{2}$ crystal form a triangular non-Lavazier sublattice with lattice vectors ${\overrightarrow{R}}_1=a\hat{x}$, ${\overrightarrow{R}}_2=a\left(\frac{\hat{x}}{2}-\frac{\sqrt{3}}{2}\hat{y}\right)$, and ${\overrightarrow{R}}_3=a(-\frac{\hat{x}}{2}-\frac{\sqrt{3}}{2})$. Here, $a=3.193\ A^0$ is the lattice distance between the two nearest neighbor Mo-atom and $\hat{x}$ and $\hat{y}$ are unit vectors in $x$- and $y$-direction, respectively.  It has been shown that,  due to the reflection symmetry in $\hat{z}$ --direction, only the hybridization of $d_{z^2}$ , $d_{xy}$, and $d_{x^2-y^2}$ happens  which opens an energy band gap at $K$- and $(-K)$- point in the first Brillouin zone (BZ) [33].  Also, it has been shown that, due to the strong spin-orbit coupling (SOC)  originated from $d_{z^2}$ --orbital, the conduction band-edge state remains spin degenerate at $K$-point whereas  the valence band-edge state splits [33]. Since, $C_{3h}$ is the group of the wave vector at band edges ($K$), one can choose $\left|\left.{\phi }_c\right\rangle =\left|\left.d_{z^2}\right\rangle \right.\right.$ and $\left|\left.{\phi }^{\tau }_v\right\rangle \right.=\frac{1}{\sqrt{2}}(\left|\left.d_{x^2-y^2}\right\rangle \right.+i\tau \left|\left.d_{xy}\right\rangle \right.)$ as the basis functions [33].  Here, $c$($v$) indicates the conduction (valence) band and $\tau =\pm 1$ is the valley degree of freedom. The $\overrightarrow{k}\cdot \overrightarrow{p}$  Hamiltonian for charge carrier (electron or hole) in the low-energy regime and near $K(-K)$-point can be written as [55]:

\begin{equation*} \label{GrindEQ__1_}
{\hat{H}}^{\tau }=\left[at\left(\tau k_x{\widehat{\sigma }}_x+k_y{\widehat{\sigma }}_y\right)
+\frac{\Delta }{2}{\widehat{\sigma }}_z\right]\otimes {\hat{I}}_s
+\tau \lambda \frac{\left({\hat{I}}_z-{\widehat{\sigma }}_z\right)}{2}\otimes {\hat{s}}_z
\end{equation*}
\begin{equation}
+{\gamma }_R\left(\tau {\widehat{\sigma }}_x\otimes {\hat{s}}_y-{\widehat{\sigma }}_y\otimes {\hat{s}}_x\right) 
\end{equation}
 
     where, $a$, $t$, $\Delta $, $\lambda $, and ${\gamma }_R$ are the lattice constant, the hooping parameter, the band gap energy, the intrinsic and Rashba SOC strength, respectively. ${\widehat{\sigma }}_i$ and ${\hat{s}}_i$ are pseudo-spin and real spin sublattice Pauli matrix, respectively, and ${\hat{I}}_i$ is $2\times 2$ unit matrix. By choosing ${\left({\psi }_{A,\uparrow },{\psi }_{A,\downarrow },{\psi }_{B,\uparrow },{\psi }_{B,\downarrow }\right)}^{\dagger }$as basis function and $\tau =+1$, the Hamiltonian reads (in $k$--space) [55]:

\begin{equation} \label{GrindEQ__2_} 
{\hat{H}}^+=\left( \begin{array}{cc}
 \begin{array}{cc}
{\Delta }/{2} & 0 \\ 
0 & {\Delta }/{2} \end{array}
 &  \begin{array}{cc}
atk_+ & 0 \\ 
2i{\gamma }_R & atk_+ \end{array}
 \\ 
 \begin{array}{cc}
atk_- & -2i{\gamma }_R \\ 
0 & atk_- \end{array}
 &  \begin{array}{cc}
\lambda -{\Delta }/{2} & 0 \\ 
0 & -\lambda -{\Delta }/{2} \end{array}
 \end{array}
\right) 
\end{equation} 
where, $k_{\pm }=k_x\pm ik_y$. 
 The nanoribbon of Mo atoms is composed by repeating a supercell which includes six Mo-atom. The Hamiltonian of the supercell reads (in real space) [33,55]:
\begin{equation} \label{GrindEQ__3_} 
H_{00}=\left( \begin{array}{ccc}
F^1_k & R &  \begin{array}{ccc}
0_{2\times 2} & 0_{2\times 2} &  \begin{array}{cc}
0_{2\times 2} & 0_{2\times 2} \end{array}
 \end{array}
 \\ 
R^{\dagger } & F^2_k &  \begin{array}{ccc}
R & 0_{2\times 2} &  \begin{array}{cc}
0_{2\times 2} & 0_{2\times 2} \end{array}
 \end{array}
 \\ 
 \begin{array}{c}
0_{2\times 2} \\ 
0_{2\times 2} \\ 
 \begin{array}{c}
0_{2\times 2} \\ 
0_{2\times 2} \end{array}
 \end{array}
 &  \begin{array}{c}
R^{\dagger } \\ 
0_{2\times 2} \\ 
 \begin{array}{c}
0_{2\times 2} \\ 
0_{2\times 2} \end{array}
 \end{array}
 &  \begin{array}{ccc}
 \begin{array}{c}
F^1_k \\ 
R^{\dagger } \\ 
 \begin{array}{c}
0_{2\times 2} \\ 
0_{2\times 2} \end{array}
 \end{array}
 &  \begin{array}{c}
R \\ 
F^2_k \\ 
 \begin{array}{c}
R^{\dagger } \\ 
0_{2\times 2} \end{array}
 \end{array}
 &  \begin{array}{cc}
 \begin{array}{c}
0_{2\times 2} \\ 
R \\ 
 \begin{array}{c}
F^1_k \\ 
R^{\dagger } \end{array}
 \end{array}
 &  \begin{array}{c}
0_{2\times 2} \\ 
0_{2\times 2} \\ 
 \begin{array}{c}
R \\ 
F^2_k \end{array}
 \end{array}
 \end{array}
 \end{array}
 \end{array}
\right) 
\end{equation} 
where, $F^1_k=\left( \begin{array}{cc}
\frac{\Delta }{2}-\mu  & 0 \\ 
0 & \frac{\Delta }{2}-\mu  \end{array}
\right)$, $F^2_k=\left( \begin{array}{cc}
\lambda -\frac{\Delta }{2}-\mu  & 0 \\ 
0 & -\lambda -\frac{\Delta }{2}-\mu  \end{array}
\right)$, and $R=\left( \begin{array}{cc}
t_0 & 0 \\ 
{2i\gamma }_R & t_0 \end{array}
\right)$. Here, $\mu $ is the chemical potential i.e., the Fermi energy at non-zero temperature [44]. Now, if we repeat the supercell m-time, a channel is formed whose Hamiltonian is as below:
\begin{equation} \label{GrindEQ__4_} 
H_C=\left( \begin{array}{ccc}
H_{00} & H_{01} &  \begin{array}{ccc}
0 & 0 &  \begin{array}{cc}
\cdots  & 0 \end{array}
 \end{array}
 \\ 
H_{10} & H_{00} &  \begin{array}{ccc}
H_{01} & 0 &  \begin{array}{cc}
\cdots  & 0 \end{array}
 \end{array}
 \\ 
 \begin{array}{c}
0 \\ 
0 \\ 
 \begin{array}{c}
\vdots  \\ 
0 \end{array}
 \end{array}
 &  \begin{array}{c}
H_{10} \\ 
0 \\ 
 \begin{array}{c}
\vdots  \\ 
0 \end{array}
 \end{array}
 &  \begin{array}{ccc}
 \begin{array}{c}
H_{00} \\ 
H_{10} \\ 
 \begin{array}{c}
\vdots  \\ 
0 \end{array}
 \end{array}
 &  \begin{array}{c}
H_{01} \\ 
H_{00} \\ 
 \begin{array}{c}
\ddots  \\ 
\cdots  \end{array}
 \end{array}
 &  \begin{array}{cc}
 \begin{array}{c}
\cdots  \\ 
\ddots  \\ 
 \begin{array}{c}
\ddots  \\ 
H_{10} \end{array}
 \end{array}
 &  \begin{array}{c}
0 \\ 
\vdots  \\ 
 \begin{array}{c}
H_{01} \\ 
H_{00} \end{array}
 \end{array}
 \end{array}
 \end{array}
 \end{array}
\right) 
\end{equation} 
Here, $H_C$ is $12m\times 12m$ matrix, $H_{10}=H^{\dagger }_{01}$ and $H_{01}$ reads:
\begin{equation} \label{GrindEQ__5_} 
H_{01}=\left( \begin{array}{ccc}
0_{2\times 2} & R &  \begin{array}{ccc}
0_{2\times 2} & 0_{2\times 2} &  \begin{array}{cc}
0_{2\times 2} & 0_{2\times 2} \end{array}
 \end{array}
 \\ 
0_{2\times 2} & 0_{2\times 2} &  \begin{array}{ccc}
0_{2\times 2} & 0_{2\times 2} &  \begin{array}{cc}
0_{2\times 2} & 0_{2\times 2} \end{array}
 \end{array}
 \\ 
 \begin{array}{c}
0_{2\times 2} \\ 
0_{2\times 2} \\ 
 \begin{array}{c}
0_{2\times 2} \\ 
0_{2\times 2} \end{array}
 \end{array}
 &  \begin{array}{c}
R \\ 
0_{2\times 2} \\ 
 \begin{array}{c}
0_{2\times 2} \\ 
0_{2\times 2} \end{array}
 \end{array}
 &  \begin{array}{ccc}
 \begin{array}{c}
0_{2\times 2} \\ 
0_{2\times 2} \\ 
 \begin{array}{c}
0_{2\times 2} \\ 
0_{2\times 2} \end{array}
 \end{array}
 &  \begin{array}{c}
R \\ 
0_{2\times 2} \\ 
 \begin{array}{c}
R \\ 
0_{2\times 2} \end{array}
 \end{array}
 &  \begin{array}{cc}
 \begin{array}{c}
0_{2\times 2} \\ 
0_{2\times 2} \\ 
 \begin{array}{c}
0_{2\times 2} \\ 
0_{2\times 2} \end{array}
 \end{array}
 &  \begin{array}{c}
0_{2\times 2} \\ 
0_{2\times 2} \\ 
 \begin{array}{c}
R \\ 
0_{2\times 2} \end{array}
 \end{array}
 \end{array}
 \end{array}
 \end{array}
\right) 
\end{equation} 
Here, $t_0=\frac{\hslash v_F}{a}=1.1$ where,  $v_F$=$0.53\times {10}^6$ m/s is the Fermi velocity in monolayer MoS${}_{2}$ [44], $\hslash =1.054\times {10}^{-34}$ m${}^{2}$Kg/s, and $a=3.193\times {10}^{10}$ m [44]. The quantum conductance versus the electron energies of the channel is calculated by using the NEGF method [58]. First, the channel is connected to the non-superconductor left ($L$) and right ($R$) semi-infinite leads and its quantum conductance ($G$) is found by using the below formula [58]:
\begin{equation} \label{GrindEQ__6_} 
G=real\left(trace\left({\mathrm{\Gamma }}_L{Gr}^r{\mathrm{\Gamma }}_R{Gr}^a\right)\right) 
\end{equation} 
where, ${\mathrm{\Gamma }}_{L(R)}$ is the  broadening matrix of left (right) lead and ${Gr}^{r(a)}$ be the retarded (advanced) Green function of channel which is:
\begin{equation} \label{GrindEQ__7_} 
{Gr}^r={\left(\left(E+i\eta \right)I-H_C-{\mathrm{\Sigma }}^r_L-{\mathrm{\Sigma }}^r_R\right)}^{-1} 
\end{equation} 
Here, $I$ is $12m\times 12m$ unit matrix, ${\mathrm{\Sigma }}^r_{L(R)}$ is the retarded self-energy of left (right) lead, and $\eta $ is an infinitesimal constant (e.g., $1.5\times {10}^{-3}$). If $g_{s,L(R)}$ is the surface Green function of left (right) lead which is calculated by suing the Sancho's method [56] then:
\begin{equation} \label{GrindEQ__8_} 
{\mathrm{\Sigma }}^r_L=\left( \begin{array}{cc}
H^{\dagger }_{01}g_{s,L}H_{01} & 0_{12\times (12m-12)} \\ 
0_{(12m-12)\times 12} & 0_{(12m-12)\times (12m-12)} \end{array}
\right) 
\end{equation} 
\begin{equation} \label{GrindEQ__9_} 
{\mathrm{\Sigma }}^R_L=\left( \begin{array}{cc}
0_{(12m-12)\times (12m-12)} & 0_{(12m-12)\times 12} \\ 
0_{12\times (12m-12)} & H_{01}g_{s,R}H^{\dagger }_{01} \end{array}
\right) 
\end{equation} 
and the broadening matrix is
\begin{equation} \label{GrindEQ__10_} 
{\mathrm{\Gamma }}_{L(R)}=i\left({\mathrm{\Sigma }}^r_{L(R)}-{\mathrm{\Sigma }}^a_{L(R)}\right) 
\end{equation} 
where, ${\mathrm{\Sigma }}^a_{L(R)}={\mathrm{\Sigma }}^{r\dagger }_{L(R)}$. 

\noindent      It should be noted that, the energy dispersion curve of the infinite nanoribbon is calculated by finding the eigenvalues of the below Hamiltonian (in $k$--space) [59]:
\begin{equation} \label{GrindEQ__11_} 
H\left(k\right)=H_{00}\left(k\right)+e^{ika}H_{01}\left(k\right)+e^{-ika}H_{10}(k) 
\end{equation} 
where, $a$ is the lattice constant of the nanoribbon.

\section{Tight-Binding model of superconductor leads}

\noindent      If  $(H_0-\mu )$ stands for the Hamiltonian of electrons, then $(\mu -TH_0T^{-1}(H^*_0(-k))$ stands for the Hamiltonian of holes where $T=i{\sigma }_yK$ is the time-reversal operator [43]. Here, $\mu $ is the chemical potential, ${\sigma }_y$ is Pauli matrix and $K$ is the complex conjugate operator. By choosing ${\left({\psi }^e_{A,\uparrow },{\psi }^h_{A,\uparrow },{\psi }^e_{A,\downarrow },{\psi }^h_{A,\downarrow },{\psi }^e_{B,\uparrow },{\psi }^h_{B,\uparrow },{\psi }^e_{B,\downarrow },{\psi }^h_{B,\downarrow }\right)}^{\dagger }$ as the basis functions and Eq.(1), the Hamiltonian of superconductor lead reads (in real space):
\begin{equation} \label{GrindEQ__12_} 
H_{SC}=\left( \begin{array}{cc}
E1 & T1 \\ 
{T1}^{\dagger } & E2 \end{array}
\right) 
\end{equation} 
where
\begin{equation} \label{GrindEQ__13_} 
E1=\left( \begin{array}{cc}
 \begin{array}{cc}
{\Delta }/{2}-\mu  & {\Delta }_{sc} \\ 
{\Delta }^{\dagger }_{sc} & -{\Delta }/{2}+\mu  \end{array}
 &  \begin{array}{cc}
0 & 0 \\ 
0 & 0 \end{array}
 \\ 
 \begin{array}{cc}
0 & 0 \\ 
0 & 0 \end{array}
 &  \begin{array}{cc}
{\Delta }/{2}-\mu  & {\Delta }_{sc} \\ 
{\Delta }^{\dagger }_{sc} & -{\Delta }/{2}+\mu  \end{array}
 \end{array}
\right) 
\end{equation} 
\begin{equation} \label{GrindEQ__14_} 
E2=\left( \begin{array}{cc}
 \begin{array}{cc}
\lambda -{\Delta }/{2}-\mu  & {\Delta }_{sc} \\ 
{\Delta }^{\dagger }_{sc} & -\lambda +{\Delta }/{2}+\mu  \end{array}
 &  \begin{array}{cc}
0 & 0 \\ 
0 & 0 \end{array}
 \\ 
 \begin{array}{cc}
0 & 0 \\ 
0 & 0 \end{array}
 &  \begin{array}{cc}
-\lambda -{\Delta }/{2}-\mu  & {\Delta }_{sc} \\ 
{\Delta }^{\dagger }_{sc} & \lambda +{\Delta }/{2}+\mu  \end{array}
 \end{array}
\right) 
\end{equation} 
\begin{equation} \label{GrindEQ__15_} 
T1=\left( \begin{array}{cc}
 \begin{array}{cc}
t_0\mathrm{\ } & 0 \\ 
0 & -t_0\mathrm{\ } \end{array}
 &  \begin{array}{cc}
0 & 0 \\ 
0 & 0 \end{array}
 \\ 
 \begin{array}{cc}
2i{\gamma }_R & 0 \\ 
0 & 2i{\gamma }_R \end{array}
 &  \begin{array}{cc}
t_0\mathrm{\ } & 0 \\ 
0 & -t_0\mathrm{\ } \end{array}
 \end{array}
\right) 
\end{equation} 
     Here, ${\Delta }_{sc}$ is the superconductor pairing potential, $t_0$ is the hopping parameter, and $A(B)$ stands for the lattice sites. It should be noted that, the above Hamiltonian refers to the spin-singlet s-wave superconductor. 

\noindent      Now, the Hamiltonian of a supercell including six Mo-atom reads:
\begin{equation} \label{GrindEQ__16_} 
H^{SC}_{00}=\left( \begin{array}{cc}
 \begin{array}{ccc}
E1 & T1 & 0_{4\times 4} \\ 
{T1}^{\dagger } & E2 & T1 \\ 
0_{4\times 4} & {T1}^{\dagger } & E1 \end{array}
 &  \begin{array}{ccc}
0_{4\times 4} & 0_{4\times 4} & 0_{4\times 4} \\ 
0_{4\times 4} & 0_{4\times 4} & 0_{4\times 4} \\ 
T1 & 0_{4\times 4} & 0_{4\times 4} \end{array}
 \\ 
 \begin{array}{ccc}
0_{4\times 4} & 0_{4\times 4} & {T1}^{\dagger } \\ 
0_{4\times 4} & 0_{4\times 4} & 0_{4\times 4} \\ 
0_{4\times 4} & 0_{4\times 4} & 0_{4\times 4} \end{array}
 &  \begin{array}{ccc}
E2 & T1 & 0_{4\times 4} \\ 
{T1}^{\dagger } & E1 & T1 \\ 
0_{4\times 4} & {T1}^{\dagger } & E2 \end{array}
 \end{array}
\right) 
\end{equation} 
   By repeating the supercell, the infinite nanoribbon is formed whose Hamiltonian is:
\begin{equation} \label{GrindEQ__17_} 
H^{Ribbon}_{SC}=\left( \begin{array}{ccc}
H^{SC}_{00} & H^{SC}_{01} &  \begin{array}{ccc}
0 & 0 &  \begin{array}{cc}
\cdots  & 0 \end{array}
 \end{array}
 \\ 
H^{SC}_{10} & H^{SC}_{00} &  \begin{array}{ccc}
H^{SC}_{01} & 0 &  \begin{array}{cc}
\cdots  & 0 \end{array}
 \end{array}
 \\ 
 \begin{array}{c}
0 \\ 
0 \\ 
 \begin{array}{c}
\vdots  \\ 
0 \end{array}
 \end{array}
 &  \begin{array}{c}
H^{SC}_{10} \\ 
0 \\ 
 \begin{array}{c}
\vdots  \\ 
0 \end{array}
 \end{array}
 &  \begin{array}{ccc}
 \begin{array}{c}
H^{SC}_{00} \\ 
H^{SC}_{10} \\ 
 \begin{array}{c}
\vdots  \\ 
0 \end{array}
 \end{array}
 &  \begin{array}{c}
H^{SC}_{01} \\ 
H^{SC}_{00} \\ 
 \begin{array}{c}
\ddots  \\ 
\cdots  \end{array}
 \end{array}
 &  \begin{array}{cc}
 \begin{array}{c}
\cdots  \\ 
\ddots  \\ 
 \begin{array}{c}
\ddots  \\ 
H^{SC}_{10} \end{array}
 \end{array}
 &  \begin{array}{c}
0 \\ 
\vdots  \\ 
 \begin{array}{c}
H^{SC}_{01} \\ 
H^{SC}_{00} \end{array}
 \end{array}
 \end{array}
 \end{array}
 \end{array}
\right) 
\end{equation} 
where, $H^{SC}_{10}=H^{SC\ \dagger }_{01}$ and $H_{01}$ reads:
\begin{equation} \label{GrindEQ__18_} 
H_{01}=\left( \begin{array}{ccc}
0_{4\times 4} & {R1}^{SC} &  \begin{array}{ccc}
0_{4\times 4} & 0_{4\times 4} &  \begin{array}{cc}
0_{4\times 4} & 0_{4\times 4} \end{array}
 \end{array}
 \\ 
0_{4\times 4} & 0_{4\times 4} &  \begin{array}{ccc}
0_{4\times 4} & 0_{4\times 4} &  \begin{array}{cc}
0_{4\times 4} & 0_{4\times 4} \end{array}
 \end{array}
 \\ 
 \begin{array}{c}
0_{4\times 4} \\ 
0_{4\times 4} \\ 
 \begin{array}{c}
0_{4\times 4} \\ 
0_{4\times 4} \end{array}
 \end{array}
 &  \begin{array}{c}
{R1}^{SC} \\ 
0_{4\times 4} \\ 
 \begin{array}{c}
0_{4\times 4} \\ 
0_{4\times 4} \end{array}
 \end{array}
 &  \begin{array}{ccc}
 \begin{array}{c}
0_{4\times 4} \\ 
0_{4\times 4} \\ 
 \begin{array}{c}
0_{4\times 4} \\ 
0_{4\times 4} \end{array}
 \end{array}
 &  \begin{array}{c}
{R1}^{SC} \\ 
0_{4\times 4} \\ 
 \begin{array}{c}
{R1}^{SC} \\ 
0_{4\times 4} \end{array}
 \end{array}
 &  \begin{array}{cc}
 \begin{array}{c}
0_{4\times 4} \\ 
0_{4\times 4} \\ 
 \begin{array}{c}
0_{4\times 4} \\ 
0_{4\times 4} \end{array}
 \end{array}
 &  \begin{array}{c}
0_{4\times 4} \\ 
0_{4\times 4} \\ 
 \begin{array}{c}
{R1}^{SC} \\ 
0_{4\times 4} \end{array}
 \end{array}
 \end{array}
 \end{array}
 \end{array}
\right) 
\end{equation} 
where,
\begin{equation} \label{GrindEQ__19_} 
{R1}^{SC}=\left( \begin{array}{cc}
 \begin{array}{cc}
t_0 & 0 \\ 
0 & -t_0 \end{array}
 &  \begin{array}{cc}
0 & 0 \\ 
0 & 0 \end{array}
 \\ 
 \begin{array}{cc}
2i{\gamma }_R & 0 \\ 
0 & 2i{\gamma }_R \end{array}
 &  \begin{array}{cc}
t_0 & 0 \\ 
0 & -t_0 \end{array}
 \end{array}
\right) 
\end{equation} 
    The quantum conductance of the nanoribbon is calculated by using the NEGF method [56] (i.e., by using Eq. (6) to Eq.(10)) and its energy dispersion is calculated by using Eq.(11). It should be noted that, the hopping parameter in $k$--space is $atk$, for both electron and hole.

\section{Tight-Binding model of topological superconductor leads}

\noindent 

 It has been shown that, the general Hamiltonian, near $\pm K$ points and in the basis ${\left(c_{e\uparrow },c_{e\downarrow }\right)}^{\dagger }$ which $c_{e\uparrow (\downarrow )}$ is the annihilation operator of $4d^2_z$ electrons with spin-up (down), reads [34]:
\begin{equation*} \label{GrindEQ__20_} 
H\left(\overrightarrow{k}+\tau \overrightarrow{K}\right)=\left(\frac{{\left|\overrightarrow{k}\right|}^2}{2m^*_e}-\mu \right)I_z+\tau \lambda {\sigma }_z+
\end{equation*}
\begin{equation}
{\gamma }_R(k_y\hat{x}-k_x\hat{y})\cdot \left({\sigma }_x\hat{x}+{\sigma }_y\hat{y}+{\sigma }_z\hat{z}\right) 
\end{equation} 
where, $I_z$ and ${\sigma }_i$ are unit and Pauli matrix, respectively and $\tau $ is the valley degree of freedom. The intrinsic and Rashba SOC parameter are $\lambda $ and ${\gamma }_R$, respectively.

\noindent        The different possible superconducting pairing phases in MoS${}_{2}$ has been studied, before [34]. It has been shown that, if $A_1$  be the trivial  representation of $C_{3h}$ point group symmetry of the monolayer of MoS${}_{2}$, the singlet pairing component of the superconducting pairing matrix dominates and , in the basis ${\left(c_{e\uparrow },c_{e\downarrow }\right)}^{\dagger }$, can be written as ${{\Delta }_s=\Delta }_0\sum^3_{j=1}{{\omega }^{j-1}{\mathrm{cos} (\overrightarrow{k}\ }.{\overrightarrow{R}}_j)}\left(i{\sigma }_y\right)$ i.e., [34]:
\begin{equation*} \label{GrindEQ__21_} 
{\Delta }_s=\left( \begin{array}{cc}
0 & {\Delta }_0\sum^3_{j=1}{{\omega }^{j-1}{\mathrm{cos} (\overrightarrow{k}\ }.{\overrightarrow{R}}_j)} \\ 
-{\Delta }_0\sum^3_{j=1}{{\omega }^{j-1}{\mathrm{cos} (\overrightarrow{k}\ }.{\overrightarrow{R}}_j)} & 0 \end{array}
\right)=
\end{equation*}
\begin{equation}
\left( \begin{array}{cc}
0 & {\Delta }_{sc} \\ 
-{\Delta }_{sc} & 0 \end{array}
\right) 
\end{equation} 
 Here, ${\Delta }_0$ and ${\sigma }_y$ are the spin-singlet pairing strength and the Pauli matrix, respectively and $\omega =e^{2\pi i/3}$  [34].  It can be shown that only this phase is characterized by the Chern numbers (CN) and supports Majorana edge states (MES) [34].

\noindent       By choosing ${\left({\psi }^e_{A,\uparrow },{\psi }^h_{A,\uparrow },{\psi }^e_{A,\downarrow },{\psi }^h_{A,\downarrow },{\psi }^e_{B,\uparrow },{\psi }^h_{B,\uparrow },{\psi }^e_{B,\downarrow },{\psi }^h_{B,\downarrow }\right)}^{\dagger }$ as the basis functions and using Eqs.(1), Eq.(20), and Eq.(21), the Hamiltonian of topological superconductor lead reads (in $k$-space):
\begin{equation} \label{GrindEQ__22_} 
H_{SC}=\left( \begin{array}{cc}
E11 & T11 \\ 
{T11}^{\dagger } & E22 \end{array}
\right) 
\end{equation} 
where
\begin{equation} \label{GrindEQ__23_} 
E11=\left( \begin{array}{cc}
 \begin{array}{cc}
\frac{k^2}{2m^*_e}-\mu  & {\Delta }_{sc} \\ 
{\Delta }^{\dagger }_{sc} & -\frac{k^2}{2m^*_h}+\mu  \end{array}
 &  \begin{array}{cc}
0 & 0 \\ 
0 & 0 \end{array}
 \\ 
 \begin{array}{cc}
0 & 0 \\ 
0 & 0 \end{array}
 &  \begin{array}{cc}
\frac{k^2}{2m^*_e}-\mu  & -{\Delta }_{sc} \\ 
-{\Delta }^{\dagger }_{sc} & -\frac{k^2}{2m^*_h}+\mu  \end{array}
 \end{array}
\right) 
\end{equation} 
\begin{equation} \label{GrindEQ__24_} 
E22=\left( \begin{array}{cc}
 \begin{array}{cc}
\lambda -\frac{k^2}{2m^*_e}-\mu  & {\Delta }_{sc} \\ 
{\Delta }^{\dagger }_{sc} & -\lambda +\frac{k^2}{2m^*_h}+\mu  \end{array}
 &  \begin{array}{cc}
0 & 0 \\ 
0 & 0 \end{array}
 \\ 
 \begin{array}{cc}
0 & 0 \\ 
0 & 0 \end{array}
 &  \begin{array}{cc}
-\lambda -\frac{k^2}{2m^*_e}-\mu  & {-\Delta }_{sc} \\ 
{-\Delta }^{\dagger }_{sc} & \lambda +\frac{k^2}{2m^*_h}+\mu  \end{array}
 \end{array}
\right) 
\end{equation} 
\begin{equation} \label{GrindEQ__25_} 
T1=\left( \begin{array}{cc}
 \begin{array}{cc}
atk & 0 \\ 
0 & atk \end{array}
 &  \begin{array}{cc}
0 & 0 \\ 
0 & 0 \end{array}
 \\ 
 \begin{array}{cc}
2i{\gamma }_R & 0 \\ 
0 & 2i{\gamma }_R \end{array}
 &  \begin{array}{cc}
atk & 0 \\ 
0 & atk \end{array}
 \end{array}
\right) 
\end{equation} 
     It should be noted that, in real space, the hopping parameter is $t_0$ ($-t_0$) for electrons (holes).

\noindent

\section{ Tight-Binding model of Josephson junction}

\noindent  Consider a superconductor-semiconductor- superconductor (SC/SM/SC) structure whose leads can be topological or non-topological superconductor. The semi-infinite superconductor leads are composed by repeating a supercell including six Mo-atom and the semiconductor flake is composed by $m$ supercells, each of them includes six Mo-atom. The quantum transport properties of the superconductor lead and the semiconductor channel are explained in the above sections II-IV. Here, we consider an effective discrete lattice description of the junction and not an atomic model [51] and study the quantum transport properties of SC/SM/SC structure by using the NEGF method (Appendix A) [51]. Since both the electron- and hole-current should be considered in the semiconductor flake, the Hamiltonian of each supercell is found by setting ${\Delta }_{sc}=0$ in Eqs.(12) to Eq.(16). By repeating the supercell $m$ times, the flake including $6m$ Mo-atom is formed. Since a $24\times 24$ matrix is considered for each supercell (Eq.16), a $24m\times 24m$ matrix should be considered for the flake. Of course, a $24\times 24$ hopping matrix between the central supercell and its right (left) neighbor supercell is considered based on the Eq.(18) (the dagger of Eq. (18)).                                                                                                 

\noindent    It has been shown that, due to the Andreev bound states, the current should has sinusoidal behavior when a phase difference ($\Delta \varphi $) is applied to the leads which changes at the range $\left(0,2\pi \right)$ (Appendix A) [51]. The Andreev bound states in the monolayer MoS${}_{2}$ has been studied, too [37,44]. It has been shown that the Andreev conductance of the superconductor MoS${}_{2}$ /normal MoS${}_{2}$ (S/N) junction, with p(n)-doped S and N regions, depends on the material properties of MoS${}_{2}$, such as the chemical potential $\left|{\mu }_N\right|$ and ${\mu }_S$ of normal and superconductor region, respectively, the intrinsic SOC strength ($\lambda $) and the superconductor pairing potential, ${\Delta }_S$ [44]. For example, for ${\mu }_S=-1.5\ eV,\ \ \lambda =0.08\ eV,\ {\Delta }_S=0.01$ eV, it has been shown that, when $\left|{\mu }_N\right|$ is smaller than 0.8 eV, the Andreev conductance is equal to zero [44]. Therefore, it is expected that its related Josephson current be equal to zero, too.  

\noindent      The detailed procedure for calculating the current operator is explained in Appendix A.   The current operator for SC/SM junction reads:
\begin{equation*} \label{GrindEQ__27_} 
I^{op}_{L(R)}=\frac{e}{h}\left[G^r\left(E\right){\mathrm{\Sigma }}^<_{L\left(R\right)}-
{\mathrm{\Sigma }}^<_{L\left(R\right)}G^a\left(E\right)\right]+
\end{equation*}
\begin{equation}
\left[G^<\left(E\right){\mathrm{\Sigma }}^a_{L\left(R\right)}-{\mathrm{\Sigma }}^r_{L\left(R\right)}G^<(E)\right]{\tau }_z 
\end{equation} 
where, $G^r$, $G^a$, and $G^<$ are the retarded, advanced and lesser Green function of SM-region, respectively. Also, ${\mathrm{\Sigma }}^r_{L\left(R\right)}$ ,${\mathrm{\Sigma }}^a_{L\left(R\right)}$,  and  ${\mathrm{\Sigma }}^<_{L\left(R\right)}$ are the retarded, advanced and lesser self-energy of left (right) SC-region, respectively. The total current is:
\begin{equation} \label{GrindEQ__28_} 
I^{\varphi }_{L(R)}=\int{dE\ real\left(trace\left(I^{op}_{L\left(R\right)}(E,\varphi )\right)\right)} 
\end{equation} 
where, $\varphi (-\varphi )$ is the applied phase to the left (right) lead at the range $(0,\pi )$ i.e., ${\Delta }_{sc,L(R)}={\Delta }_0e^{i\varphi (-\varphi )}$. It should be noted that, here,
\begin{equation} \label{GrindEQ__29_} 
{\tau }_z={\sigma }_z\otimes {\mathrm{l}}_{N\times N}={\left( \begin{array}{ccc}
1 & 0 &  \begin{array}{ccc}
0 & \cdots  & 0 \end{array}
 \\ 
0 & -1 &  \begin{array}{ccc}
0 & \cdots  & 0 \end{array}
 \\ 
 \begin{array}{c}
\vdots  \\ 
0 \\ 
0 \end{array}
 &  \begin{array}{c}
\vdots  \\ 
0 \\ 
0 \end{array}
 &  \begin{array}{ccc}
 \begin{array}{c}
\ddots  \\ 
\cdots  \\ 
\cdots  \end{array}
 &  \begin{array}{c}
\vdots  \\ 
1 \\ 
0 \end{array}
 &  \begin{array}{c}
\vdots  \\ 
0 \\ 
-1 \end{array}
 \end{array}
 \end{array}
\right)}_{24m\times 24m} 
\end{equation} 
  and, ${\mathrm{\Sigma }}^r_L$ and ${\mathrm{\Sigma }}^r_{R\ }$are:
\begin{equation} \label{GrindEQ__30_} 
{\mathrm{\Sigma }}^r_L=\left( \begin{array}{cc}
{\beta }^{\dagger }g_{sL}\beta {\times 10}^{-3} & 0_{24\times (24m-24)} \\ 
0_{(24m-24)\times 24} & 0_{(24m-24m)\times (24m-24)} \end{array}
\right) 
\end{equation} 
\begin{equation} \label{GrindEQ__31_} 
{\mathrm{\Sigma }}^r_R=\left( \begin{array}{cc}
0_{(24m-24)\times (24m-24)} & 0_{(24m-24)\times 24} \\ 
0_{24\times (24m-24)} & \beta g_{sR}{\beta }^{\dagger }\times {10}^{-3} \end{array}
\right) 
\end{equation} 
where, $\beta $-matrix is
\begin{equation} \label{GrindEQ__32_} 
\beta =\left( \begin{array}{cc}
 \begin{array}{ccc}
0_{4\times 4} & T & 0_{4\times 4} \\ 
0_{4\times 4} & 0_{4\times 4} & 0_{4\times 4} \\ 
0_{4\times 4} & T & 0_{4\times 4} \end{array}
 &  \begin{array}{ccc}
0_{4\times 4} & 0_{4\times 4} & 0_{4\times 4} \\ 
0_{4\times 4} & 0_{4\times 4} & 0_{4\times 4} \\ 
T & 0_{4\times 4} & 0_{4\times 4} \end{array}
 \\ 
 \begin{array}{ccc}
0_{4\times 4} & 0_{4\times 4} & 0_{4\times 4} \\ 
0_{4\times 4} & 0_{4\times 4} & 0_{4\times 4} \\ 
0_{4\times 4} & 0_{4\times 4} & 0_{4\times 4} \end{array}
 &  \begin{array}{ccc}
0_{4\times 4} & 0_{4\times 4} & 0_{4\times 4} \\ 
T & 0_{4\times 4} & T \\ 
0_{4\times 4} & 0_{4\times 4} & 0_{4\times 4} \end{array}
 \end{array}
\right) 
\end{equation} 
where,
\begin{equation} \label{GrindEQ__33_} 
T=\left( \begin{array}{cc}
 \begin{array}{cc}
t_0 & 0 \\ 
0 & {-t}_0 \end{array}
 &  \begin{array}{cc}
0 & 0 \\ 
0 & 0 \end{array}
 \\ 
 \begin{array}{cc}
2i{\gamma }_R & 0 \\ 
0 & 2i{\gamma }_R \end{array}
 &  \begin{array}{cc}
t_0 & 0 \\ 
0 & {-t}_0 \end{array}
 \end{array}
\right) 
\end{equation}

\noindent       Also, the real-valued singularities of the density of states (DOS) are the Andreev bound state (ABS) energies and are computed as a function of the phase difference of the order parameter of the leads [50]. For each $\Delta \varphi $, the total DOS is (Appendix A):
\begin{equation} \label{GrindEQ__34_} 
DOS=\sum_E{DOS\left(E\right)}=\sum_E{\left(\frac{1}{2\pi }trace\left[A\left(E\right)\right]
=\frac{1}{2\pi }trace(i\left\{G^r-G^a\right\})\mathrm{\ }\right)} 
\end{equation} 
 It should be noted that for observing the Andreev bound states, we should set  the energy of carriers (holes and electrons) at range $-{\Delta }_0\times {10}^{-3}\le E\le {\Delta }_0\times {10}^{-3}$ [43].

\section{  Results and Discussion}

\subsection{\textbf{ Semiconductor Channel}}

Here, we set $a=$3.193 $A^0$,  $\lambda =0.08$ eV, ${\gamma }_R=0.033$ $eV$,  $\Delta =1.9$ eV, $t=1.1$ eV, $m^*_e=0.5m_e$, and $\mu =0.83$ eV [43]. Fig.1(a) shows the energy dispersion curve of the nanoribbon for two values of chemical potentials i.e., $\mu =0\ $and $\ 0.83$ eV. The bad gap energy is equal to 1.82 eV ($\Delta -\lambda $) (for spin-up electrons) and $1.98$ eV ($\Delta +\lambda $) for (spin-down) at $k=0$. The spin splitting is equal to $0.16$ eV ($2\lambda $), in the valence band but the conduction band is spin degenerate. Also, Fig.1(b) shows the quantum conductance versus electron energies. As it shows, the transport gap is non-zero and, in consequence, the infinite nanoribbon is a direct band gap semiconductor.

\noindent       Fig.2 shows the quantum conductance versus electron energies of a flake including 360 Mo-atom which is formed by repeating a supercell including six Mo-atom. As the figure shows, in comparison with the infinite nanoribbon, the flake habits as a semiconductor, too.

\begin{figure}
\centering{}
\includegraphics[width=.8\linewidth]{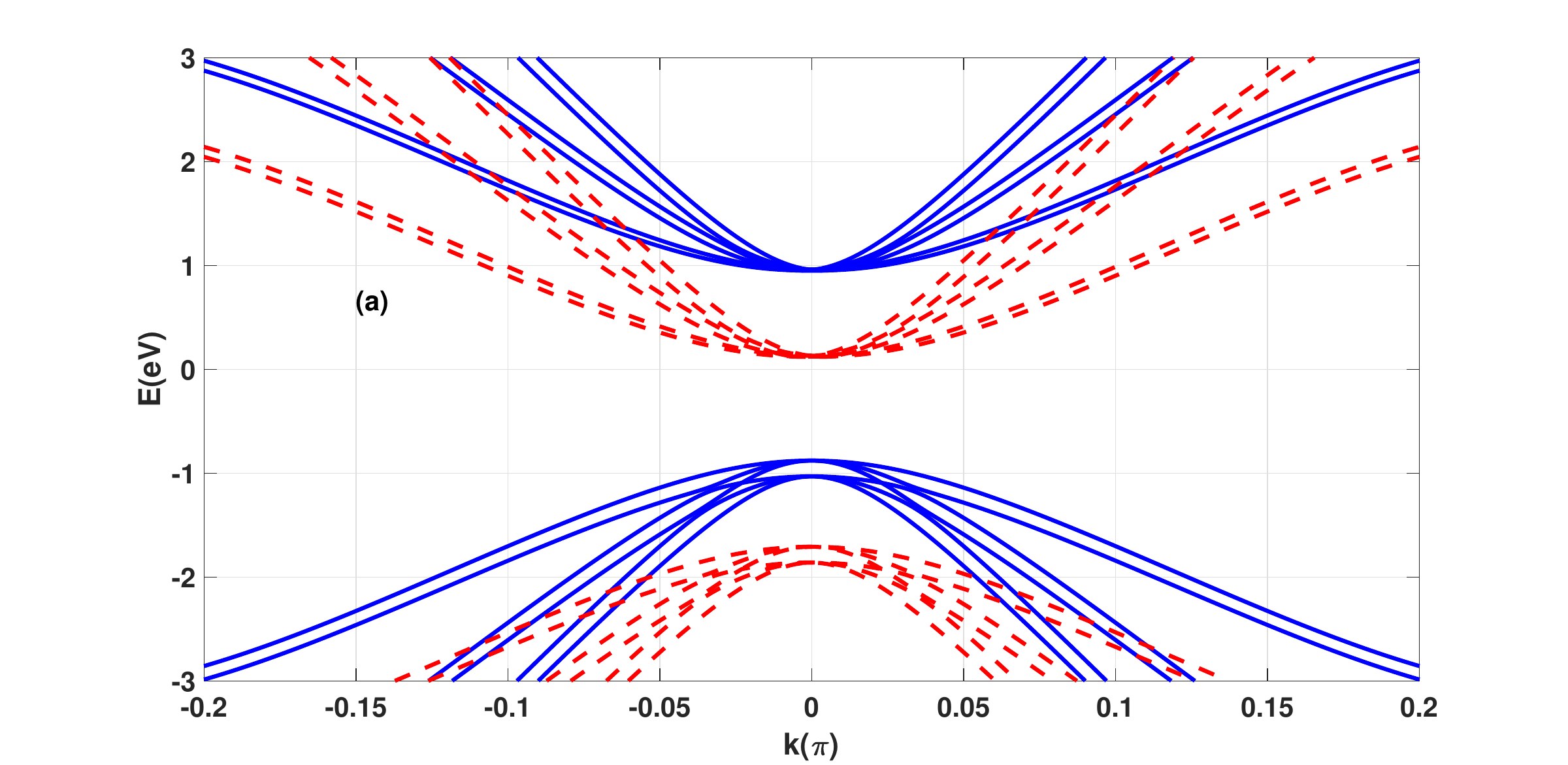}
\includegraphics[width=.8\linewidth]{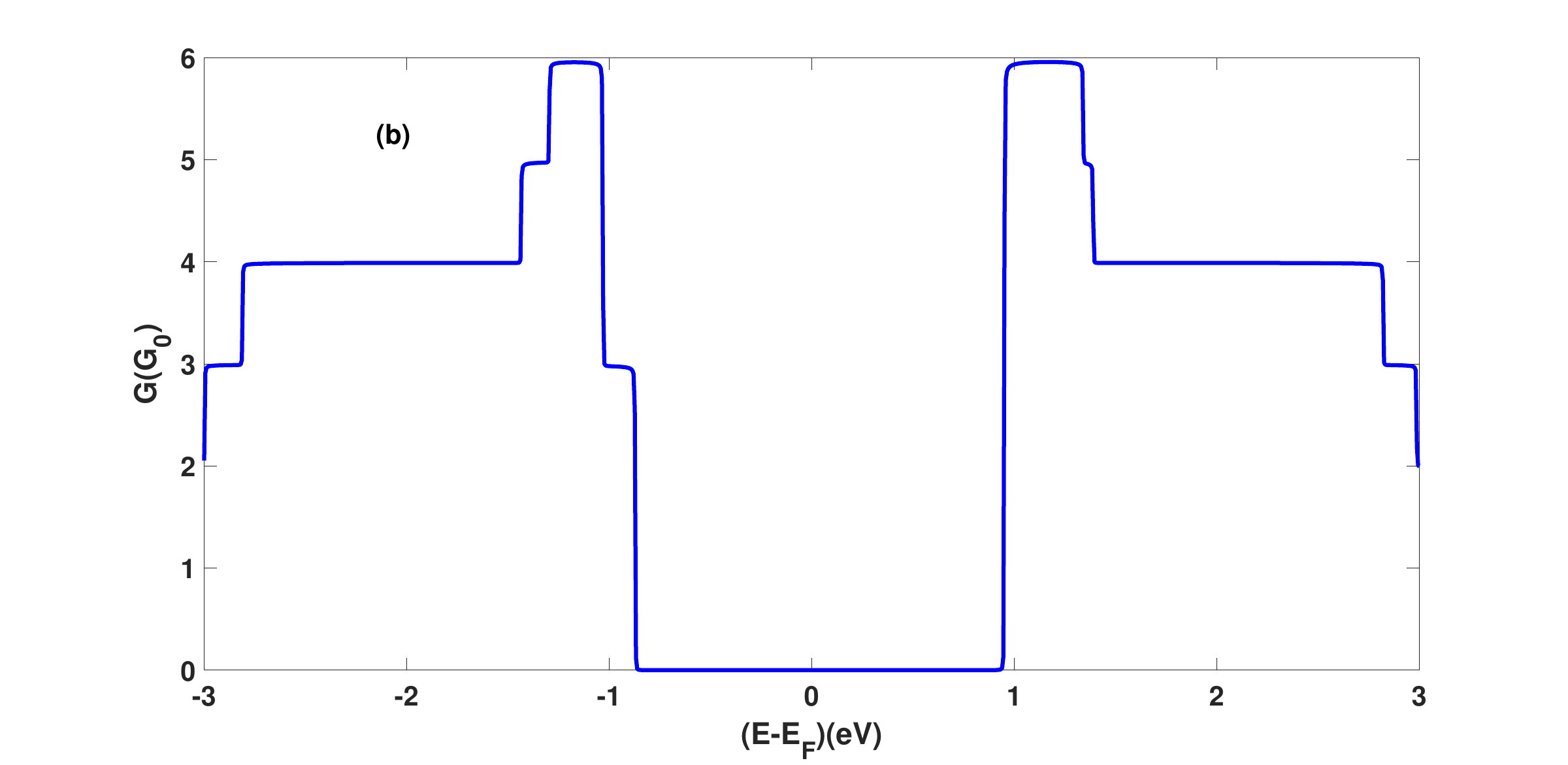}
\caption{\textbf{(Color online) (a) Energy dispersion of an infinite zigzag nanoribbon composed by repeating a supercell including six Mo-atom. The curve with}}$\boldsymbol{\mu }\boldsymbol{=}\boldsymbol{0}\boldsymbol{\ (}\boldsymbol{0}.\boldsymbol{83}\boldsymbol{)}$\textbf{ eV is shown in filled (dashed) blue (red) color. (b) The quantum conductance vs electron energies with non-zero transport gap. The infinite nanoribbon habits as a direct band gap semiconductor.}
\textbf{\label{fig:{Fig1}}}
\end{figure}   

\begin{figure}
\centering{}
\includegraphics[width=.8\linewidth]{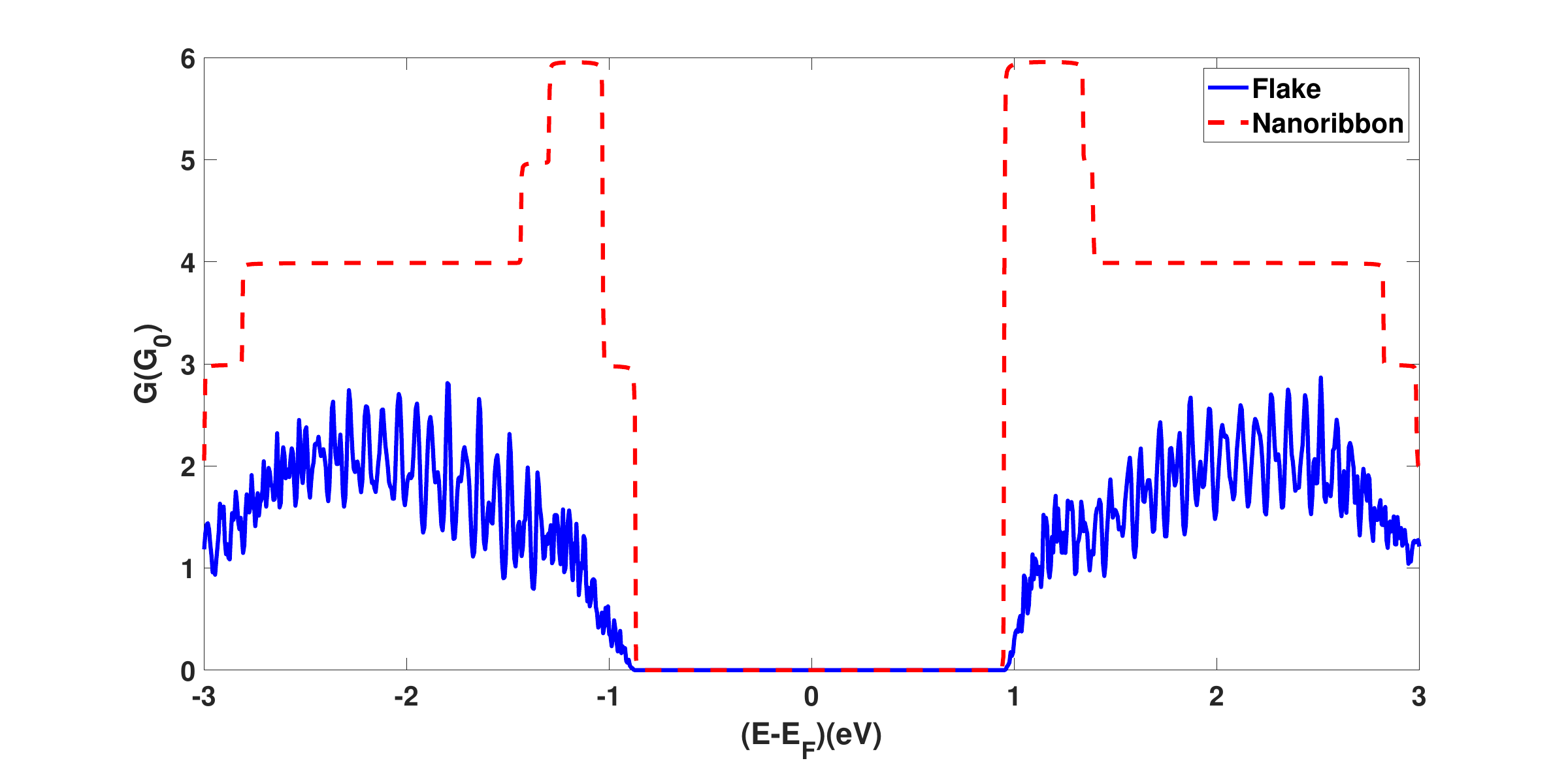}
\caption{\textbf{(Color online) The quantum conductance versus electron energies of flake including 360 Mo-atom which is formed by repeating a supercell including six Mo-atom. The curve of flake (infinite nanoribbon) is shown in filled blue color (dashed red color). In comparison with the infinite nanoribbon, the flake habits as a semiconductor.}}
\textbf{\label{fig:{Fig2}}}
\end{figure}   

\noindent 

\subsection{\textbf{Superconductor lead}}

 Here, we set $a=$3.193 $A^0$,  $\lambda =0.08$ eV, ${\gamma }_R=0.033$ $eV$,  $\Delta =1.9$ eV, $t=1.1$ eV, $m^*_e=0.5m_e$, ${\Delta }_{SC}=0.01$ eV and $\mu =0$, $0.83$ eV [43]. By attention to the basis function i.e.,  ${\left({\psi }^e_{A,\uparrow },{\psi }^h_{A,\uparrow },{\psi }^e_{A,\downarrow },{\psi }^h_{A,\downarrow },{\psi }^e_{B,\uparrow },{\psi }^h_{B,\uparrow },{\psi }^e_{B,\downarrow },{\psi }^h_{B,\downarrow }\right)}^{\dagger }$, and Eqs.(13) to (16), the bottom of electron conduction band ($k=0$) is composed by spin-up and spin-down electrons $({\Delta }/{2}-\mu =0.95-\mu $) and the bottom of hole valence band is composed by sin-up $(-\lambda +{\Delta }/{2}+\mu =0.875+\mu )$ and spin-down $(\lambda +{\Delta }/{2}+\mu =1.025+\mu )$ holes. Therefore, hole states are spin degenerate but electron states are not in the region $E>0$. Also, the top of electron valence band ($k=0$) is composed by spin-up ($\lambda -{\Delta }/{2}-\mu =-0.875-\mu $) and spin-down ($-\lambda -{\Delta }/{2}-\mu =-1.025-\mu $) electrons and the top of the hole conduction band is composed by the spin-up and spin-down ($-{\Delta }/{2}+\mu =-0.95+\mu $) holes. Therefore, electron states are spin degenerate but hole states are not in the region $E<0$. Fig.3 (a) shows the carrier (electrons plus holes) energy dispersion of the superconductor infinite nanoribbon for both values of  $\mu =0$ and $0.83$ eV. For $\mu =0$ eV and in $E>0$ region, the filled blue curve includes three values $0.95$ eV ( due to the electrons), $0.875$ eV and $1.025$ eV (both due to the holes) at $k=0$. For $\mu =0.83$ eV and in $E>0$ region, the dashed red curve includes three values $0.12$ eV (due to the electrons), $1.705$ eV and $1.855$ eV (both due to the holes) at $k=0$. Also, for $\mu =0$ eV and in $E<0$ region, the filled blue curve includes three values $-0.95$ eV (due to the holes), $-0.875$ eV and $-1.025$ eV (both due to the electrons) at $k=0$. For $\mu =0.83$ eV and in $E<0$ region, the dashed red curve includes three values $-0.12$ eV (due to the holes), $-1.705$ eV and $-1.855$ eV (both due to the electrons) at $k=0$. Also, Fig.3 (b) shows the band splitting at $E<0$ region for $\mu =0.83$ eV. The energy band gap is $E^{e,h\uparrow }_g=\Delta -\lambda =1.82$ eV and $E^{e,h\downarrow }_g=\Delta +\lambda =1.98$ eV when the electrons and holes are considered alone, for both values of the chemical potentials. 

\noindent       Fig.4 shows the quantum conductance versus carrier (electrons plus holes) energies of superconductor infinite nanoribbon. Since, both electrons and holes take part in the quantum transport, the transport band gap depends on the transport of both of them. As Fig.3(a) shows, for $\mu =0\ $ ($0.83$), the difference between the bottom of filled blue (dashed red) curve at $E>0$ region and its top at $E<0$ is equal to $1.75$ ($0.24$) eV. As Fig.4 shows, the transport band gap is equal to $1.75$ ($0.24$) eV for $\mu =0 (0.83)$ eV, approximately, which is in good agreement with the above mentioned differences. It means that, the transport band gap decreases by increasing the chemical potential due to the contribution of both electrons and holes in the quantum transport. Also, it is not equal to the energy band gap of electrons (holes) alone.

\noindent       It should be noted that the carriers whose energy is less than the superconductor pairing potential (i.e., $E<{\Delta }_s$ ) form the Cooper pairs and do not take part in the quantum transport. Since, ${\Delta }_s=0.01$ eV,  and the transport gap is $1.75$ eV ($\mu =0\ eV$) and $0.24$ eV ($\mu =0.83\ eV$), the non-zero quantum transport is attributed to the carriers with energy $\left|E\right|>1.75\left(0.24\right)\ eV>{\Delta }_s$. 

 \begin{figure}
\centering{}
\includegraphics[width=.8\linewidth]{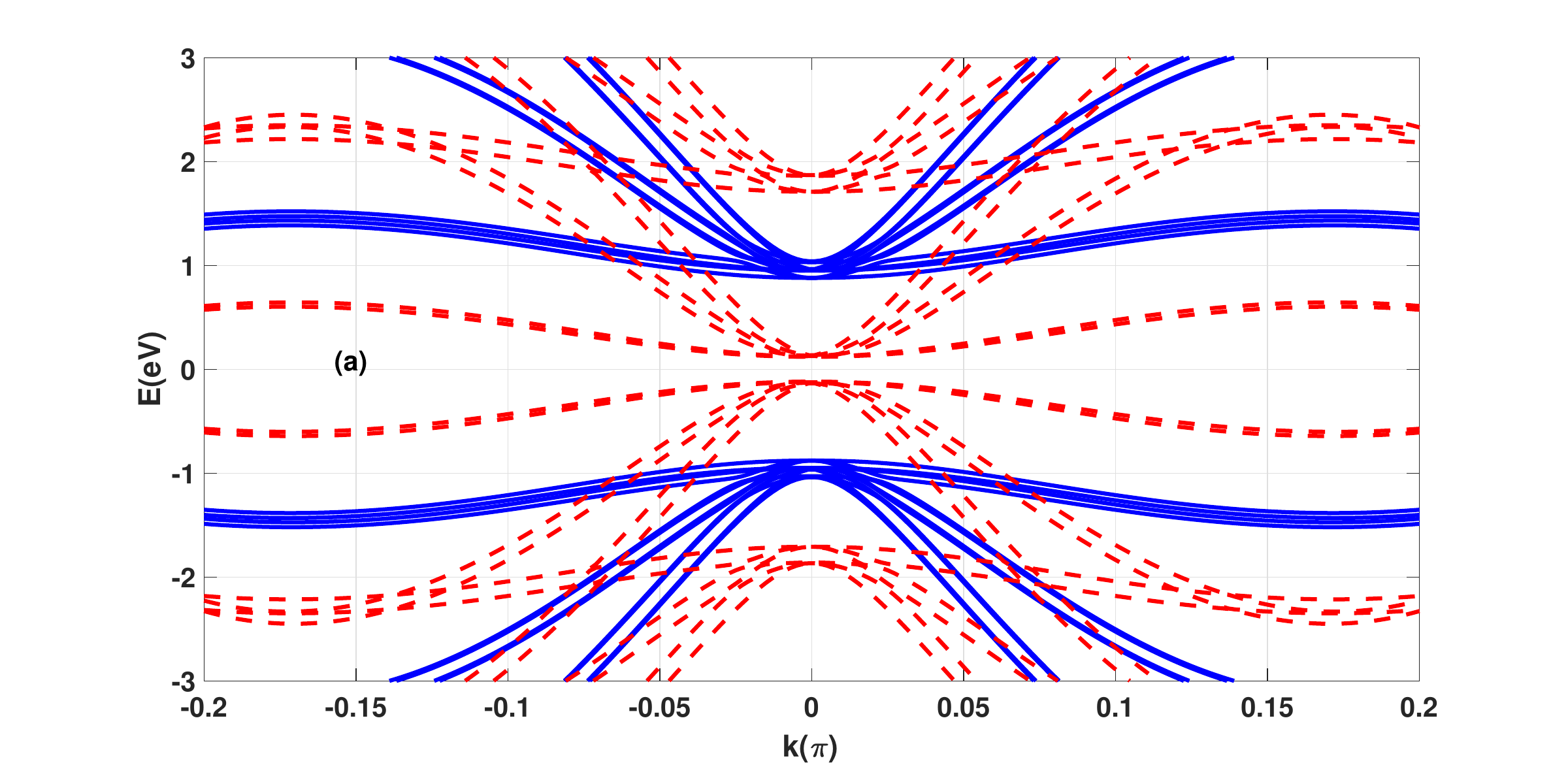}
\includegraphics[width=.8\linewidth]{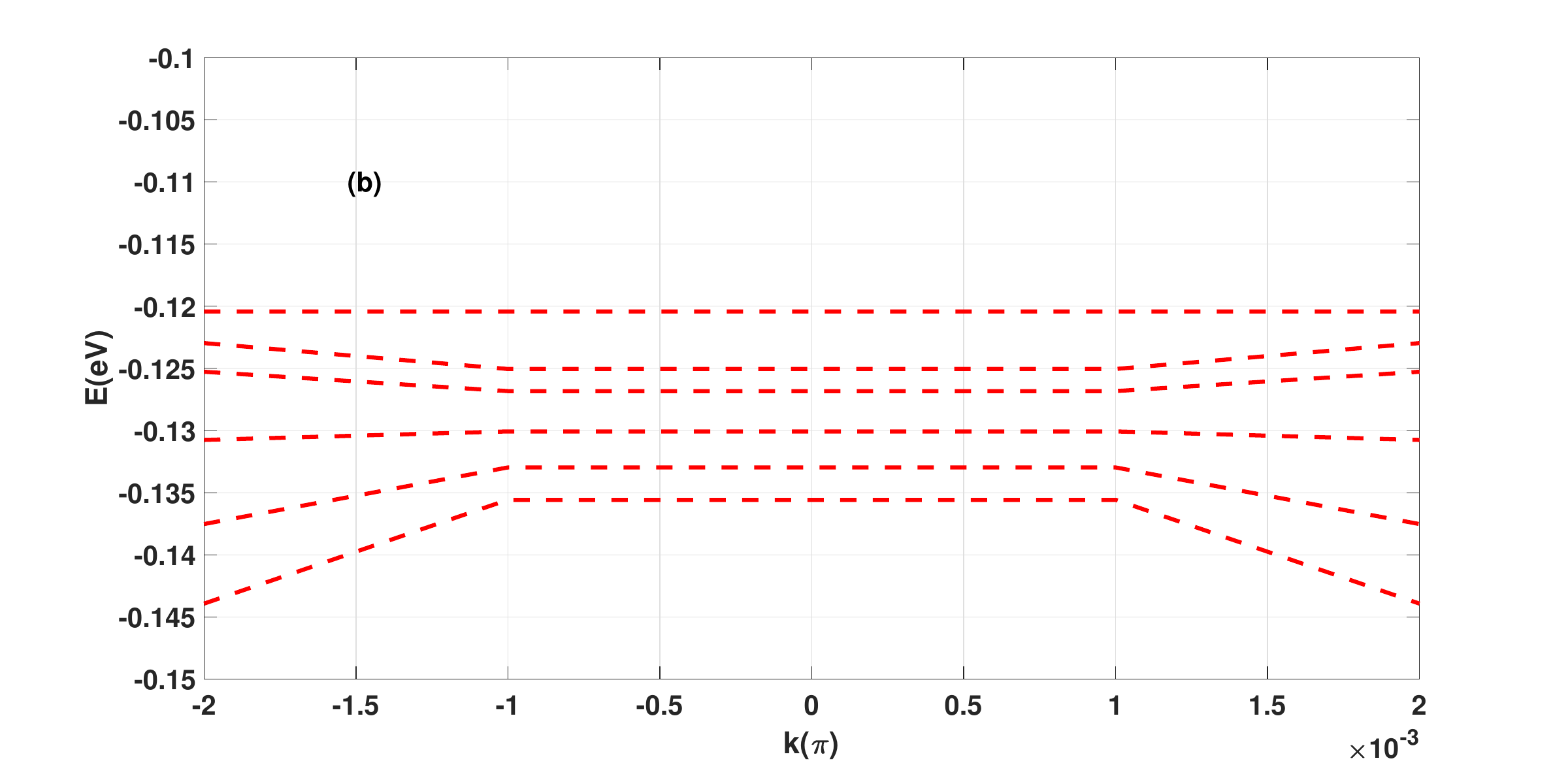}
\caption{\textbf{(Color online) The energy dispersion curve of superconductor lead. (a) The curve with $\boldsymbol{\mu }\boldsymbol{=}\boldsymbol{0}$\textbf{ eV (}$\boldsymbol{0}.\boldsymbol{83}\boldsymbol{\ }$\textbf{eV) is shown in filled (dashed) blue (red) color. (b) Non-degenerate curves in }$\boldsymbol{E}\boldsymbol{<}\boldsymbol{0}$\textbf{ region when }$\boldsymbol{\mu }\boldsymbol{=}\boldsymbol{0}.\boldsymbol{83}$\textbf{ eV. }}}
\textbf{\label{fig:{Fig3}}}
\end{figure}  

\begin{figure}
\centering{}
\includegraphics[width=.8\linewidth]{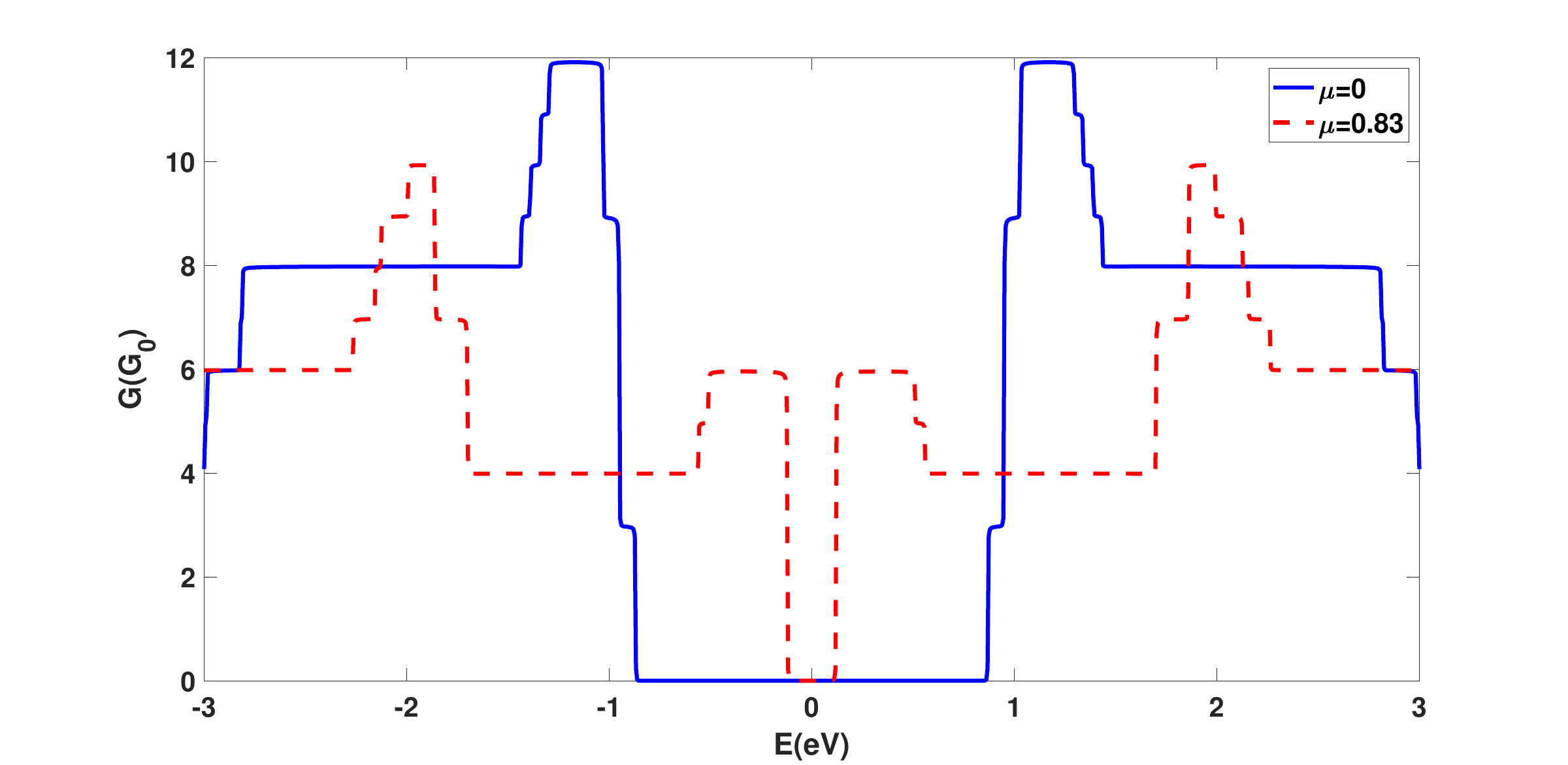}
\caption{\textbf{(Color online)The carrier (electrons plus holes) quantum conductance curve of superconductor infinite nanoribbon for $\boldsymbol{\mu }\boldsymbol{=}\boldsymbol{0}\boldsymbol{\ (}\boldsymbol{0}.\boldsymbol{83}\boldsymbol{)}$\textbf{ ev in filled (dashed) blue (red) color. The non-zero quantum transport is attributed to the carriers with energy }$\left|\boldsymbol{E}\right|\boldsymbol{\mathrm{>}}\boldsymbol{\mathrm{1}}.\boldsymbol{\mathrm{75}}\left(\boldsymbol{\mathrm{0}}.\boldsymbol{\mathrm{24}}\right)\boldsymbol{\mathrm{\ }}\boldsymbol{eV}\boldsymbol{\mathrm{>}}{\boldsymbol{\mathrm{\Delta }}}_{\boldsymbol{s}}$\textbf{.}}}
\textbf{\label{fig:{Fig4}}}
\end{figure}  

\subsection{\textbf{ Topological superconductor leads}}

 Here, we set $a=$3.193 $A^0$,  $\lambda =0.08$ eV, ${\gamma }_R=0.033$ $eV$,  $\Delta =1.9$ eV, $t=1.1$ eV, $m^*_e=0.5m_e$, ${\Delta }_0=0.01$ eV and $\mu =0$, $0.83$ eV [44]. Fig.5 shows the energy dispersion curve of carriers (electron plus holes) for three different values of $\frac{\mu }{\lambda }=0,\ 0.19,\ \ 10.1$. Yuan et al., have shown that, for $\left|\mu \right|\ll \left|\lambda \right|$, the topological superconductivity, with CN=2, can be seen in MoS${}_{2}$ [34]. As Figs. 5(a) and (b) show, for both values of  $\frac{\mu }{\lambda }=0,\ $and $0.19$, the energy gap is closed at $k=0$ and $k=0.025$, respectively, when the topological pairing potential (${\Delta }_0\sum^3_{j=1}{{\omega }^{j-1}{\mathrm{cos} (\overrightarrow{k}\ }.{\overrightarrow{R}}_j)}$) is substituted with the non-topological one (${\Delta }_0$). But, as Fig. 5(c) shows, for $\frac{\mu }{\lambda }=10.1$, there is no difference between topological and non-topological cases and in both of them band gap closing is seen at $k=0.019$. 

\noindent       Figs. 6(a), (b) and (c) show the quantum transport versus carrier (electron plus holes) energy for three different values of $\frac{\mu }{\lambda }=0,\ 0.19,\ \ 10.1$ at $k=0,\ 0.025$, $0.019$, respectively. Due to the energy gap closing for both values of  $\frac{\mu }{\lambda }=0$ and $,\ 0.19$, the non-zero conductance is seen at $E=E_F$, in topological case, which is in good agreement with the results of Figs. 5 (a) and (b). That is, The transport band gap is closed in topological cases i.e., when $\left|\mu \right|\ll \left|\lambda \right|$ i.e.\textbf{, }cases in Fig. 6(a) and (b). These non-zero conductance are attributed to the Majorana zero mode [34]. Therefore, our results are in good agreement with the results of Ref. [34] and our model can be used for explaining the topological superconductivity in MoS${}_{2}$ with CN=2.

\begin{figure}
\centering{}
\includegraphics[width=.8\linewidth]{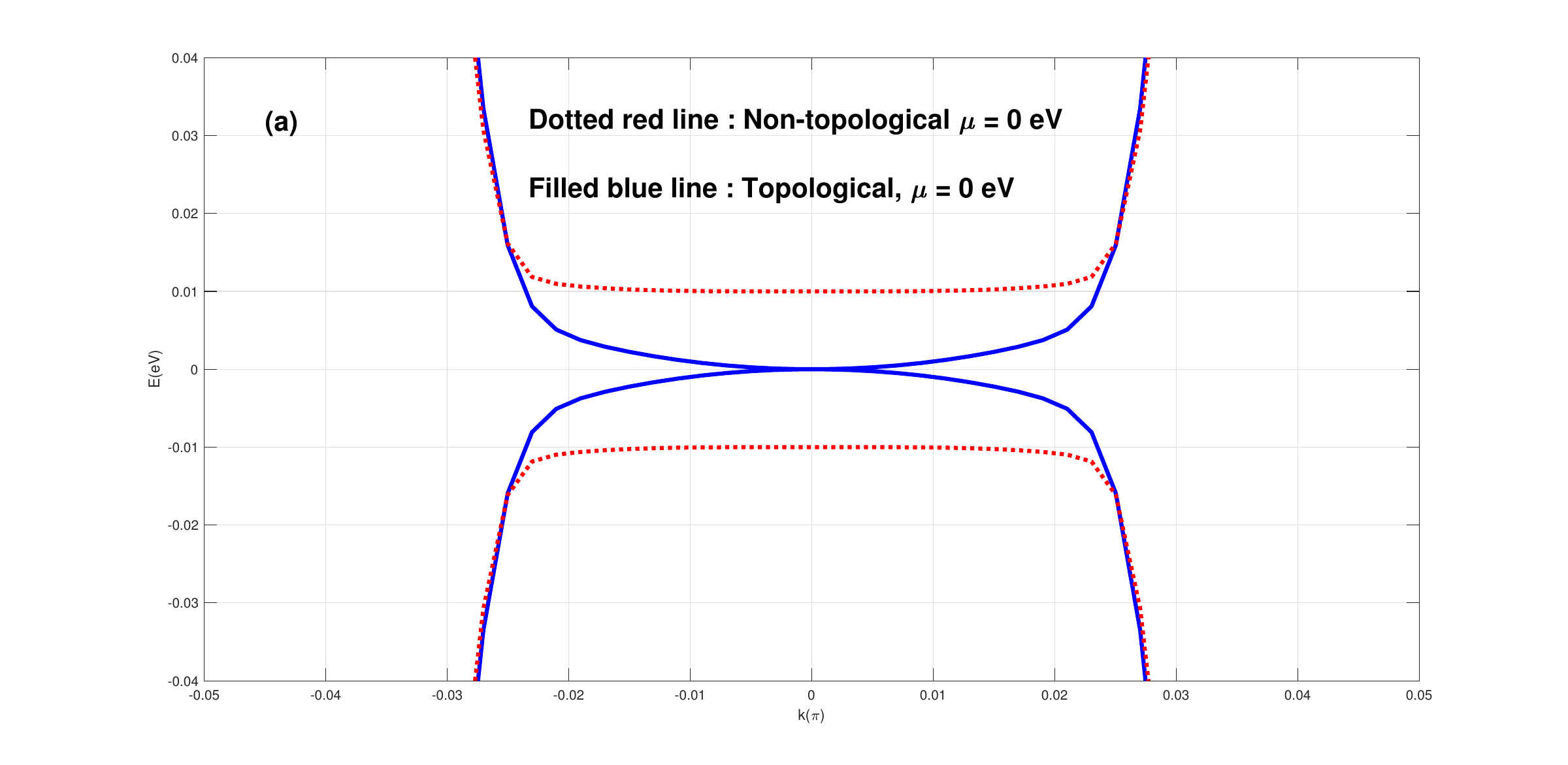}
\includegraphics[width=.8\linewidth]{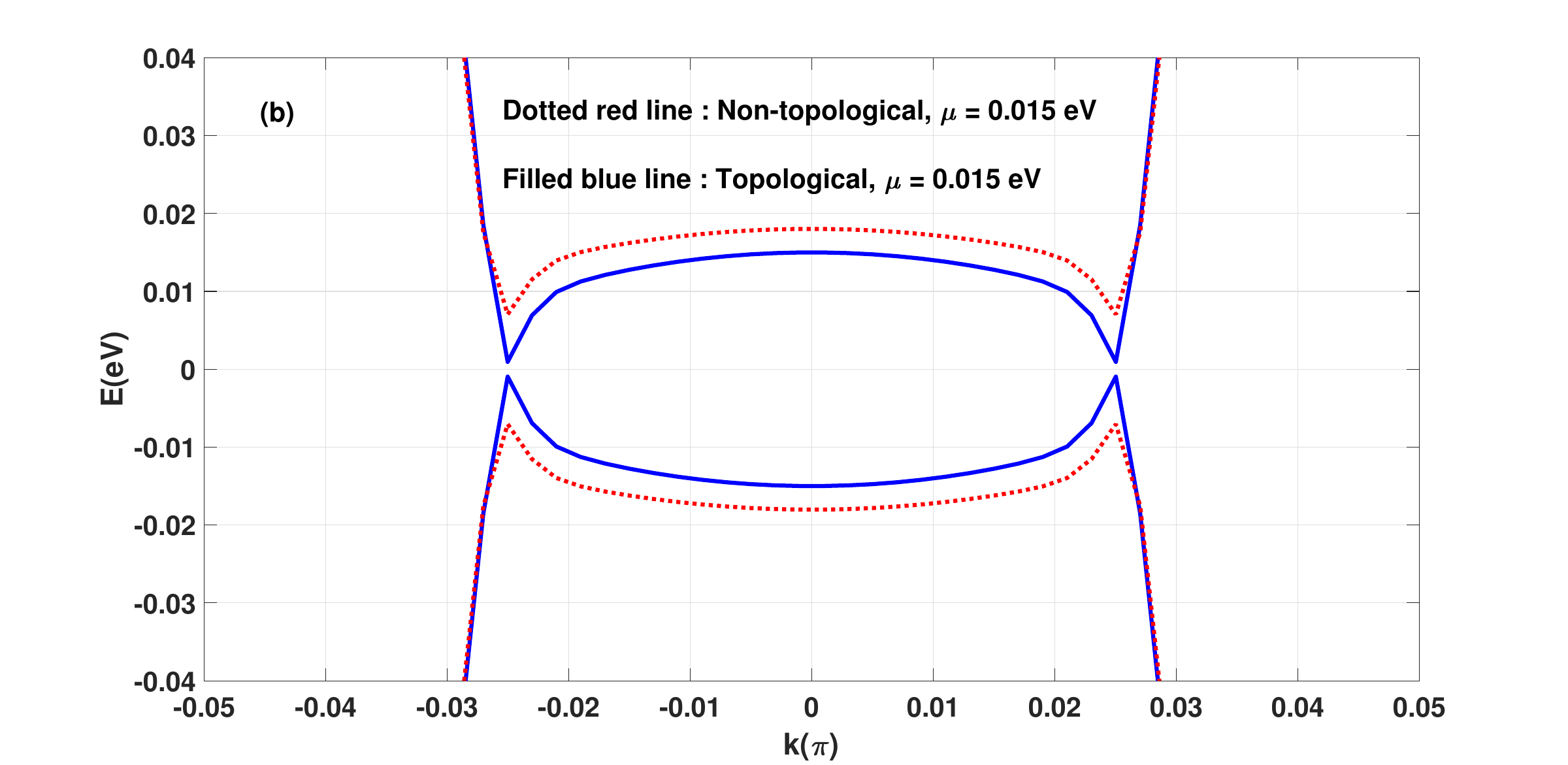}
\includegraphics[width=.8\linewidth]{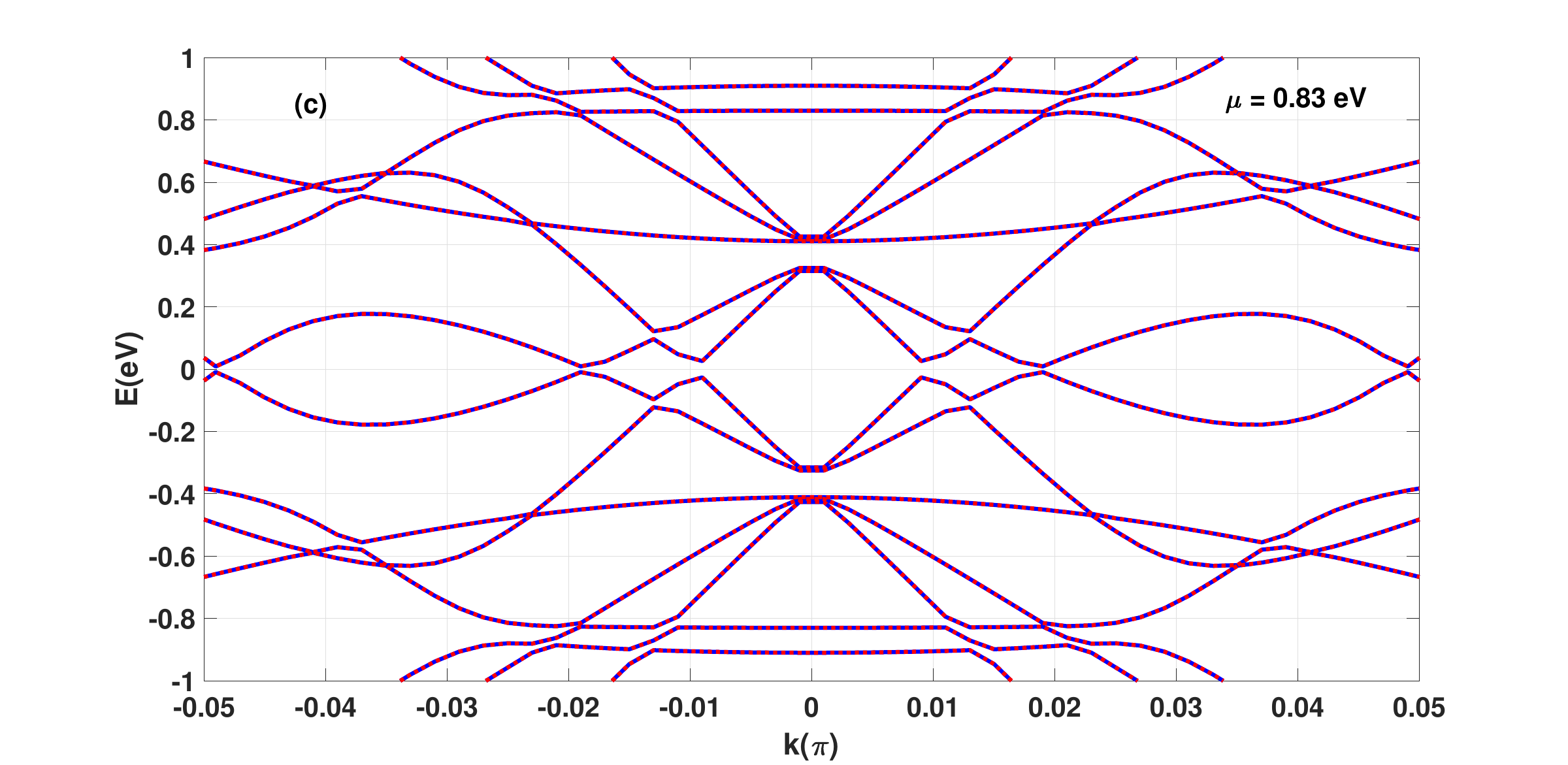}
\caption{\textbf{(Color online) The energy dispersion curve of topological (non-topological) superconducting leads for three different values of (a) $\boldsymbol{\mu }\boldsymbol{=}\boldsymbol{0}$, (b)$\boldsymbol{\mu }\boldsymbol{=}\boldsymbol{0.015}$, and (c)$\boldsymbol{\mu }\boldsymbol{=}\boldsymbol{0.83}$\textbf{ ev in filled (dashed) blue (red) color.}}}
\textbf{\label{fig:{Fig5}}}
\end{figure} 
 
\begin{figure}
\centering{}
\includegraphics[width=.8\linewidth]{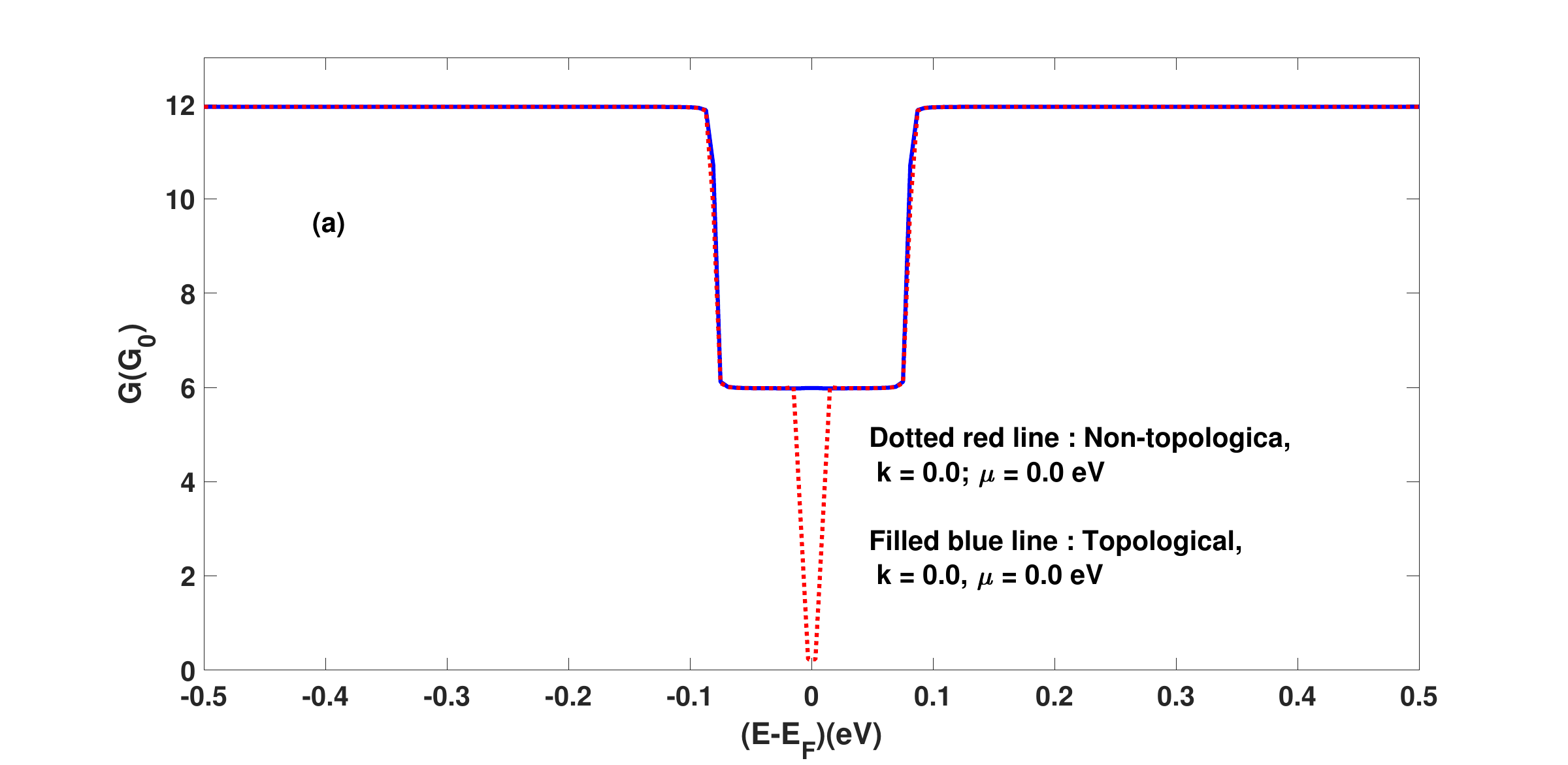}
\includegraphics[width=.8\linewidth]{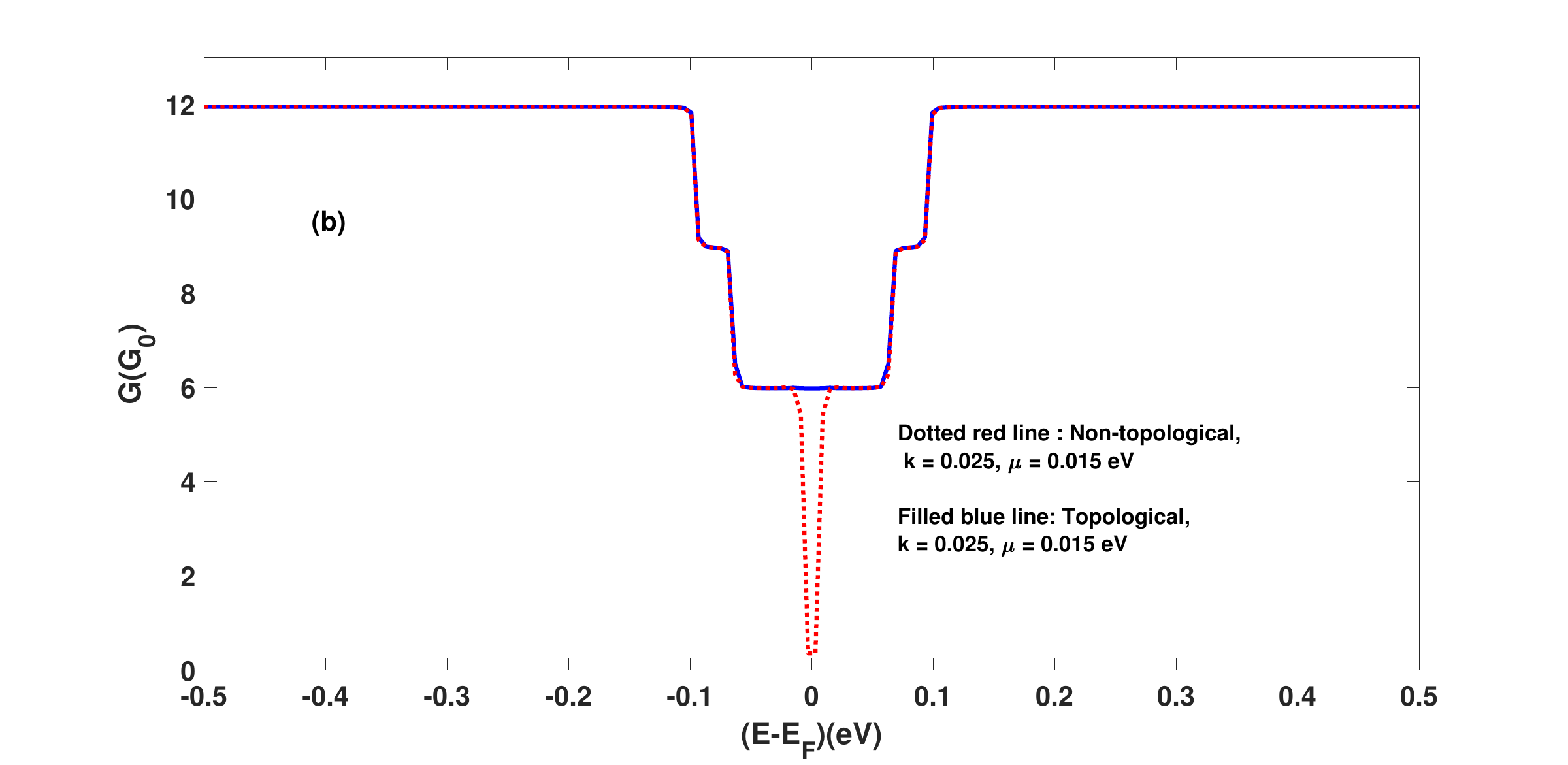}
\includegraphics[width=.8\linewidth]{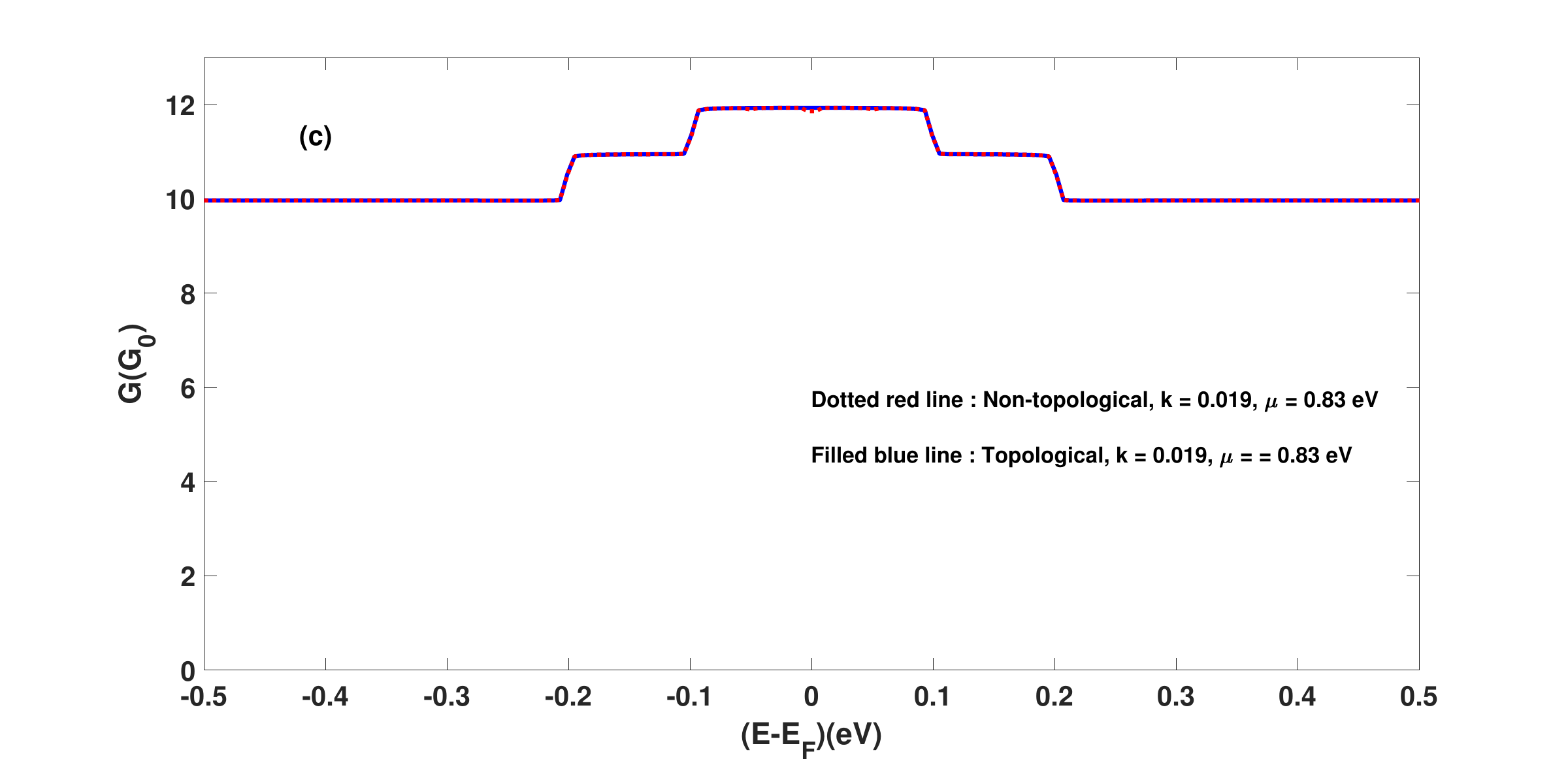}
\caption{\textbf{(Color online)  The quantum conductance versus carrier energy of topological (non-topological) superconducting leads in filled (dotted) blue (red) line for three different values of  (a) $\boldsymbol{\mu=0 }$, (a) $\boldsymbol{\mu=0.015 }$, (a) $\boldsymbol{\mu=0.83 eV }$ }}
\textbf{\label{fig:{Fig6}}}
\end{figure} 

\subsection{\textbf{ Josephson junction}}

 Let us to consider a two-dimensional monolayer MoS${}_{2}$-based Josephson junction which is composed by an intermediate semiconductor flake and the left and right semi-infinite topological and non-topological superconductor leads.  The flake includes 72 Mo-atom which is composed by repeating a supercell including six Mo-atom and the leads are composed by repeating a supercell including six Mo-atom.

\noindent            \textbf{A. Non-topological leads}

\noindent     For calculating the total Josephson current ($I_t=I_L-I_R$) and total density of states (DOS) versus the applied phase difference between the leads, we set,  $\Delta =1.9$, eV, $\lambda =0.08$ eV, $t_0=1.1$ eV (for hopping inside a supercell and between the supercells), ${\gamma }_R=0.033$ eV, $\mu =0.83$ eV, ${\mathrm{\Delta }}_0=0.01$ eV [44], $t=t_0\times {10}^{-3}$ eV (for coupling the leads to the flake), and the energy of carriers at the range $-{\Delta }_0\le E\le {\Delta }_0$ [50]. Fig. 7 (a) and (b) show $I_t$ and DOS, respectively. As Fig.7(a) shows, the total current ($I_L-I_R$) behaves as, $I_0{\mathrm{sin} (\Delta \varphi )\ }$ when the phase $\varphi $ ($-\varphi $) is applied to the left (right) lead at the range $\left(0,\pi \right)$. The behavior is attributed to the Andreev bound states and shows that he transmission value $\tau <1$ (i.e., anti-crossing regime) [48,56]. Also, as the Fig.7(b) shows, DOS has Gaussian curve due to the ABS [48-56].

\begin{figure}
\centering{}
\includegraphics[width=.8\linewidth]{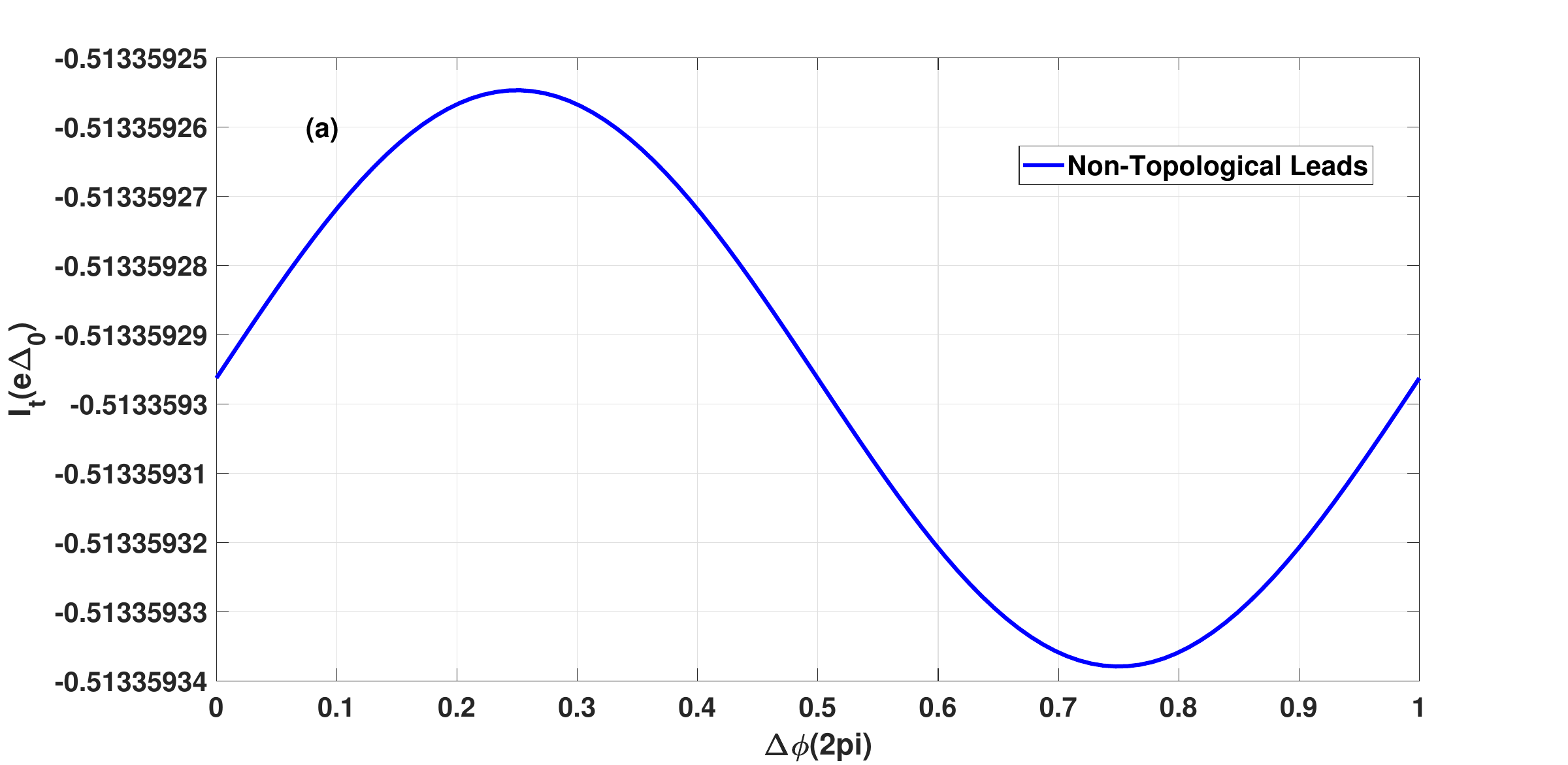}
\includegraphics[width=.8\linewidth]{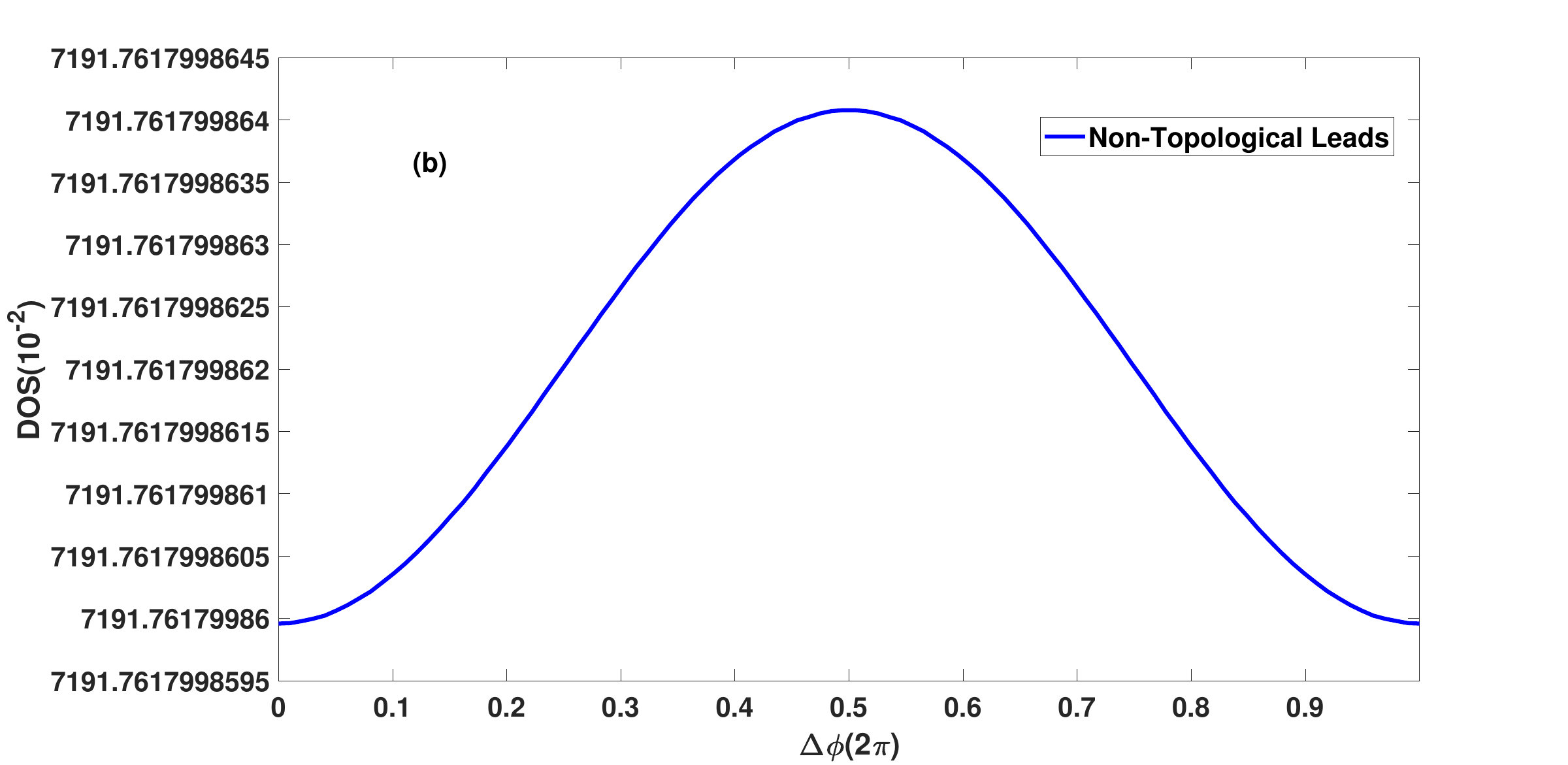}
\caption{\textbf{(Color online)  (Color online) (a) The total Josephson current ( ${\boldsymbol{I}}_{\boldsymbol{t}}\boldsymbol{=}{\boldsymbol{I}}_{\boldsymbol{L}}\boldsymbol{-}{\boldsymbol{I}}_{\boldsymbol{R}}$\textbf{) and (b) the total density of states versus the phase difference between leads.}}}
\textbf{\label{fig:{Fig7}}}
\end{figure}  

\noindent            \textbf{B. Topological leads}

As it is explained in sub-section A of section III, for observing the Majorana zero mode (MZM) the chemical potential should be smaller than the superconductor pairing potential i.e., $\left|\mu \right|<{\Delta }_0$. In this reason, we set $\mu =0$ eV and $\mu =0$.015 eV and ${\Delta }_0=0.01$ eV [44]. Fig.8 shows the total density of states versus applied phase difference to the topological superconductor leads, for these two values of $\mu $. As it shows, although DOS increases by increasing the chemical potential, but for each value of $\mu $, it is constant. Therefore, the Andreev bound states are not formed and the junction behaves as an open circuit ($I_t=0$) when $-{\Delta }_0\times {10}^{-3}\le E\le {\Delta }_0\times {10}^{-3}$. This is the expense we should pay for forming the MZM. It should be noted that, due to the Fig.5a (Fig.5b), we set $k=0$ and $0.025$ in Eqs.(23) and Eq.(24) for calculating the Josephson current and DOS when the leads are topological superconductors when $\mu =0$ and $0.015$ eV, respectivel. Also the term $atk$ is substituted by $t_0$ ($-t_0$) for electrons (holes) in Eq.(25). 

\begin{figure}
\centering{}
\includegraphics[width=.8\linewidth]{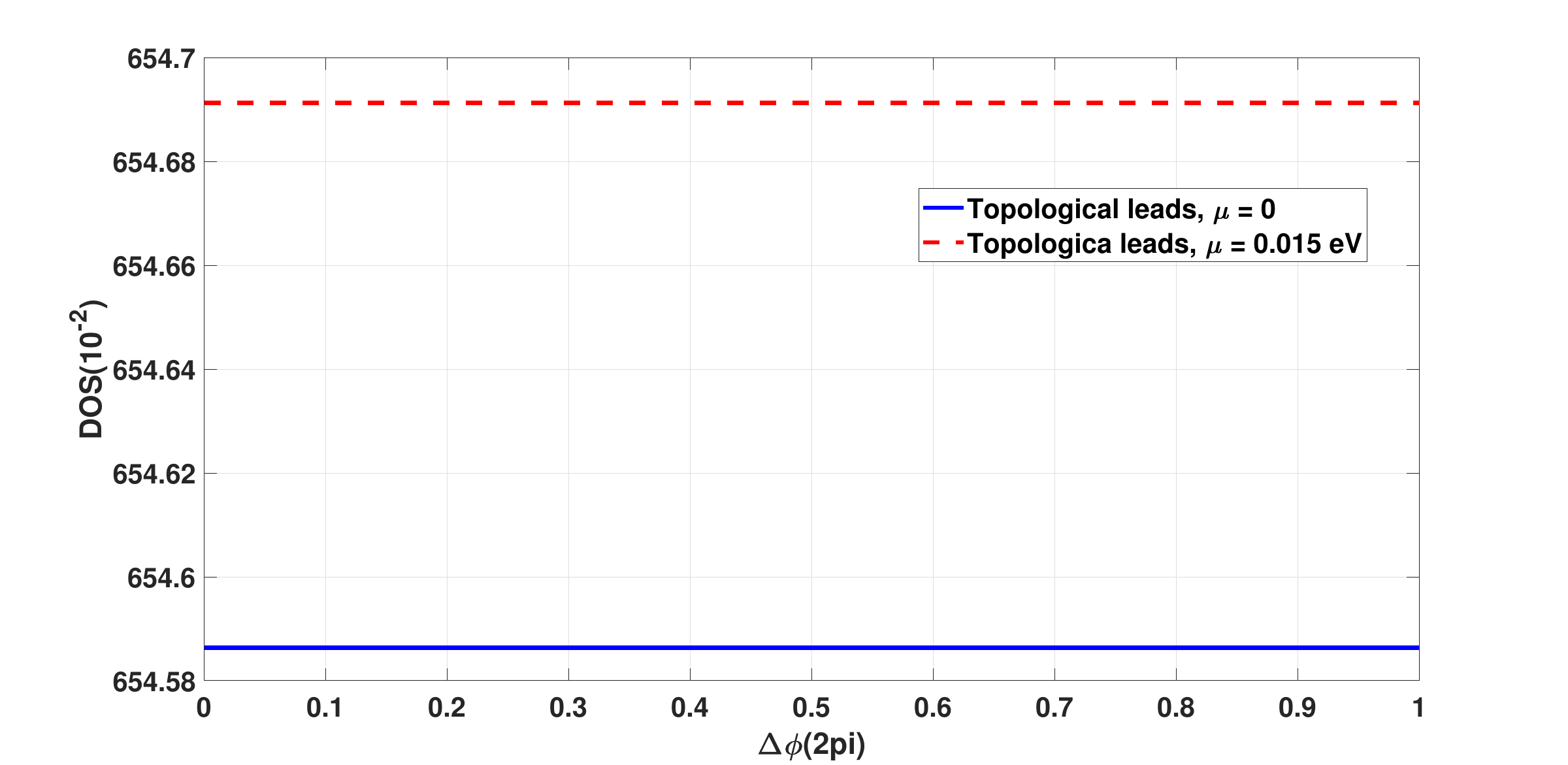}
\caption{\textbf{(Color online) The total density of states (electrons plus holes) versus applied phase difference to the topological leads in Josephson junction, when $\boldsymbol{\mu }\boldsymbol{=}\boldsymbol{0}\boldsymbol{\ (}\boldsymbol{0}.\boldsymbol{015}\boldsymbol{)}$\textbf{ eV.}}}
\textbf{\label{fig:{Fig8}}}
\end{figure} 

  Therefore, one can consider the two-dimensional monolayer MoS${}_{2}$-based Josephson junction as a two-state switch which is open when $\mu \ge 0.8$ eV and is closed when $\mu <0.8$ or $\mu =0$ (for topological leads). 

\section{Conclusion}

  Nowadays, the fault-tolerant quantum computing and computers are a hot research filed. One of the new chosen strategy for realizing the fault-tolerant is based on using the Anyons instead of Fermions (electrons) and Bosons (photons). The solid-state Majorana is an Anyon whose state function belongs to electron and hole (anti-particle of electron) simultaneously. The zero energy Majorana state (called Majorana zero mode(MZM) has the main role in developing the fault-tolerant quantum computing and computers based on the topological properties of materials. 

\noindent        By considering the non-Lavazier sublattice of Mo atoms in the infinite nanoribbon of MoS${}_{2}$, we have studied the quantum transport properties of the two-dimensional MoS${}_{2}$-based Josephson junction, with topological and non-topological (ordinary) superconductor leads. We have shown that, if the chemical potential is much greater than the superconductor pairing potential and be smaller than the hopping integral in the infinite non-topological (ordinary) superconductor MoS${}_{2}$-nanoribbon, the Andreev bound states (ABS) are formed and in consequence the Josephson current (density of states(DOS)) has sinusoidal (Gaussian) curve. Also, by assuming a special formula for superconductor pairing potential, it has been shown that, the Majorana zero modes (MZM) exist in the infinite topological superconductor nanoribbon. Then, by substituting the ordinary superconductor lead by topological lead, it has been shown that the ABS are not formed, and in consequence, its related Josephson current is zero. Therefore, the Josephson junction can be considered as a two-state switch. The open-state is attributed to the ordinary superconductor leads while the close-state is referred to the topological leads. It could be done by adjusting the chemical potential. Therefore, ABS cannot mimic the Majorana state, as zero-bias conductance. 

\noindent        Therefore, based on these results and in agreement with the global approach to developing the hybrid two-dimensional material (2DM)-CMOS technology [52], it can be concluded that the use of two-dimensional material-based Josephson junctions may be put on the research roadmap for designing and manufacturing the topological switch which is required for manufacturing the topological computers.  

\noindent

{\bf .~ Appendix A}
\renewcommand{\thefigure}{A-\arabic{figure}}
\setcounter{figure}{0} 
\renewcommand{\theequation}{A.\arabic{equation}}
\setcounter{equation}{0} 
\noindent 
{\bf }

     Consider one-dimensional normal (N) region which is connected to the semi-infinite one-dimensional superconductor (SC) leads from left and right. The Hamiltonian of normal and superconductor regions are as below [51]:

\noindent 
\begin{equation}
H_{N(SC)}=\left( \begin{array}{ccc}
{\alpha }_{N(SC)} & \beta  &  \begin{array}{ccc}
0 & \ \ \ \ \cdots \ \ \ \  & 0 \end{array}
 \\ 
{\beta }^{\dagger } & {\alpha }_{N(SC)} &  \begin{array}{ccc}
\beta \ \ \ \ \  & 0 & \ \ \ \ \ 0 \end{array}
 \\ 
 \begin{array}{c}
0 \\ 
\vdots  \\ 
0 \end{array}
 &  \begin{array}{c}
{\beta }^{\dagger } \\ 
0 \\ 
\cdots  \end{array}
 &  \begin{array}{ccc}
 \begin{array}{c}
{\alpha }_{N(SC)} \\ 
\ddots  \\ 
\cdots  \end{array}
 &  \begin{array}{c}
\beta  \\ 
\ddots  \\ 
{\beta }^{\dagger } \end{array}
 &  \begin{array}{c}
\vdots  \\ 
\beta  \\ 
{\alpha }_{N(SC)} \end{array}
 \end{array}
 \end{array}
\right)                                       
\end{equation}
\noindent 

\noindent which,  ${\alpha }_N=\left( \begin{array}{cc}
2t-\mu  & 0 \\ 
0 & -2t+\mu  \end{array}
\right)$, ${\alpha }_{SC}=\left( \begin{array}{cc}
2t-\mu  & {\Delta }_{SC} \\ 
{\Delta }^{\dagger }_{SC} & -2t+\mu  \end{array}
\right)$, $\beta =\left( \begin{array}{cc}
-t & 0 \\ 
0 & t \end{array}
\right)$ , and $t=\frac{{\hslash }^2}{2m^*a^2}$ [51]. Here,$\ \ \mu $ is the chemical potential,$\ \ t$ is the hopping constant, $\ m^*$ is the effective mass of electron, $a$ is the lattice constant, and  ${\Delta }_{SC}={\Delta }_0e^{i\varphi }$ is the superconductor pairing potential which $\varphi $ is the applied phase to the lead. It should be noted that this is an effective discrete lattice description of the junction and is not an atomic model [51]. The basis set is $\psi ={\left(c_{i,\uparrow },c^{\dagger }_{i+1,\downarrow }\right)}^{\dagger }\ $ which $c^{\dagger }_{i,\uparrow }$ and $c_{i+1,\downarrow }$ are the creation operator of spin-up electron and spin-up hole, respectively. It should be noted that the annihilation operator of spin-down electron is the creation operator of spin-up hole. 

\noindent       By using the Sancho's method [56], the retarded surface Green function of the left ($g_{sL}$) and right ($g_{sL}$) leads can be calculated. For doing that, one should consider ${\alpha }_{SC}$ as the Hamiltonian of the central cell and $\beta $ as the hopping term between the central cell and its left and right neighbor cells. In this effective discrete lattice, the retarded self-energies of leads are [51]:
\begin{equation}
{\mathrm{\Sigma }}^r_L=\left( \begin{array}{cc}
{(\beta }^{\dagger }g_{sL}\beta )\times {10}^{-3} & 0_{2\times (N-2)} \\ 
0_{(N-2)\times 2} & 0_{(N-2)\times (N-2)} \end{array}
\right)                                                                      
\end{equation}
\begin{equation}
{\mathrm{\Sigma }}^r_R=\left( \begin{array}{cc}
0_{(N-2)\times (N-2)} & 0_{(N-2)\times 2} \\ 
0_{2\times (N-2)} & (\beta g_{sR}{\beta }^{\dagger })\times {10}^{-3} \end{array}
\right)
\end{equation}                                                                       

\noindent      It should be noted that, for observing the Andreev bound states, we should consider the potential barrier between the superconductor leads and the transport channel. This is modeled by decreasing the hopping parameter, t,  i.e., by multiplying the ${\mathrm{\Sigma }}^r_{L(R)}$-matrix in the factor ${10}^{-3}$.

\noindent In coherent regime, the retarded Green function of the transport channel is (i.e., normal region) is equal to [46-51]:

\begin{equation}
G^r\left(E\right)={\left(\left(E+i\eta \right)\mathrm{l}-H_N-{\mathrm{\Sigma }}^r_L-{\mathrm{\Sigma }}^r_R\right)}^{-1}                                                                        
\end{equation}

\noindent where, $\mathrm{l}$ is $\left(N\times N\right)\ $unit matrix ($N$ is the dimension of $H_N$), ${\mathrm{\Sigma }}^r_{L(R)}$ is The self-energy of left (right) lead, and $\eta $ is an infinitesimal constant (e.g. $1.5\times {10}^{-3}$). The advanced Green function is $G^a\left(E\right)=G^{r\dagger }(E)$. Also, the lesser self-energy or in-scattering function (matrix) of either contact is:

\begin{equation}
{\mathrm{\Sigma }}^<_{L\left(R\right)}\left(E\right)=-\left({\mathrm{\Sigma }}^r_{L\left(R\right)}-{\mathrm{\Sigma }}^a_{L\left(R\right)}\right)f_{L\left(R\right)}(E)=i{\mathrm{\Gamma }}_{L\left(R\right)}f_{L\left(R\right)}(E)                                                  
\end{equation}

\noindent where, ${\mathrm{\Sigma }}^a_{L(R)}={\mathrm{\Sigma }}^{r\dagger }_{L(R)}$ is advanced self-energy and $f_{L(R)}=1/(1+{\mathrm{exp} \left(\frac{E-\mu }{K_BT}\right)\ })$ represents the Fermi-Dirac distribution function at left (right) contact. The broadening matrix of the left (right) lead is ${\mathrm{\Gamma }}_{L\left(R\right)}$. It can be shown that the current operator for left (right) contact is [47-52]:

\begin{equation}
I^{op}_{L(R)}=\frac{e}{h}\left[G^r\left(E\right){\mathrm{\Sigma }}^<_{L\left(R\right)}-{\mathrm{\Sigma }}^<_{L\left(R\right)}G^a\left(E\right)+G^<\left(E\right){\mathrm{\Sigma }}^a_{L\left(R\right)}-{\mathrm{\Sigma }}^r_{L\left(R\right)}G^<(E)\right]                                
\end{equation}

\noindent Where, $G^<\left(E\right)=G^r(E)\left({\mathrm{\Sigma }}^<_L+{\mathrm{\Sigma }}^<_R\right)G^a(E)$ is the lesser Green function. Since both contacts are superconductors, the net current through left(right) contact is equal to the difference between electron and hole current. Therefore, the fledged current operator is:

\begin{equation}
I^{op}_{L(R)}=\frac{e}{h}\left[G^r\left(E\right){\mathrm{\Sigma }}^<_{L\left(R\right)}-{\mathrm{\Sigma }}^<_{L\left(R\right)}G^a\left(E\right)+G^<\left(E\right){\mathrm{\Sigma }}^a_{L\left(R\right)}-{\mathrm{\Sigma }}^r_{L\left(R\right)}G^<(E)\right]{\tau }_z                     
\end{equation}       

\noindent 

\noindent where, by attention to the basis set which has been used for writing $H_{SC}$

\begin{equation}
{\tau }_z={\mathrm{l}}_{N\times N}\otimes {\sigma }_z={\left( \begin{array}{ccc}
1 & 0 &  \begin{array}{cc}
\cdots  & 0 \end{array}
 \\ 
0 & -1 &  \begin{array}{cc}
\cdots  & 0 \end{array}
 \\ 
 \begin{array}{c}
\vdots  \\ 
0 \end{array}
 &  \begin{array}{c}
0 \\ 
0 \end{array}
 &  \begin{array}{cc}
 \begin{array}{c}
\ddots  \\ 
\cdots  \end{array}
 &  \begin{array}{c}
\vdots  \\ 
-1 \end{array}
 \end{array}
 \end{array}
\right)}_{2N\times 2N}
\end{equation}                                                 

\noindent Now, the total current reads:

\begin{equation}
I^{\varphi }_{L(R)}=\int{dE\ real\left(trace\left(I^{op}_{L\left(R\right)}(E,\varphi )\right)\right)}                                                        
\end{equation}
\noindent By changing  the applied phase to left (right) lead at range $\left(-\pi -\pi \right)$ and integrating  $I^{\varphi }_{L(R)}$ over the phase differences , it is possible to obtain the variation of current against the variation of the applied phase differences, i.e.,

\begin{equation}
I_{L(R)}=\int{d\varphi }\ I^{\varphi }_{L(R)}                                                                             \end{equation}

\noindent       It should be noted that, when we deal with the superconducting leads (or even superconductor transport channel), we cannot write the lesser Green function as, $G^<=iG^r\left({\mathrm{\Gamma }}_LF_L+{\mathrm{\Gamma }}_RF_R\right)G^a$, where ${\mathrm{\Gamma }}_{L(R)}=i\left({\mathrm{\Sigma }}^r_{L(R)}-{\mathrm{\Sigma }}^a_{L(R)}\right)$ and $F_{L(R)}=\left( \begin{array}{cc}
f(E,\mu ) & 0 \\ 
0 & f(E,-\mu ) \end{array}
\right)$, because it misses a term in the current proportional to $\left(G^rG^a\right)$ , the trace of which increases with the number of Andreev bound states (which depends on the length of the nanowire) [47-52]. In consequence, we can write the current operator as below [47-52]:

\begin{equation}
I^{op}_{L(R)}=\frac{e}{h}real\left(trace\left(\left(G^a{\mathrm{\Sigma }}^a_{L(R)}-G^r{\mathrm{\Sigma }}^r_{L(R)}\right){\tau }_z\right)\right)                                         
\end{equation}

\noindent    Also, for each $\Delta \varphi $, the density of states (DOS) can be computed as the trace of the spectral Green function i.e., [51]:

\begin{equation}
DOS=\sum_E{DOS\left(E\right)}=\sum_E{\left(\frac{1}{2\pi }trace\left[A\left(E\right)\right]=\frac{1}{2\pi }trace(i\left\{G^r-G^a\right\})\mathrm{\ }\right)}                       
\end{equation}

\noindent 

\noindent The real-valued singularities of the density of states are the Andreev bound state (ABS) energies and are computed as a function of the phase difference of the order parameter of the leads [51]. 

\noindent       However, for observing the Andreev bound states in a typical experiment, one should set $\mu \gg {\Delta }_0$ and $E\ll {\Delta }_0$, where $E$ is the excitation energy of electrons [48,56]. Also, the parabolic dispersion relations, in normal region, correspond to the condition $t\gg \mu $ [48,56].  In consequence, the choice of the effective lattice parameter $a$ is bounded by the condition $t\gg \mu \gg {\Delta }_0$. Otherwise, Andreev bound states will not be observed [48,56]. We choose $t=1.2$ eV, $\mu =0.4$ eV and ${\Delta }_0$=0.01 eV, such that $\frac{t}{\mu }=30$ and $\frac{\mu }{{\Delta }_0}=40$. Also, we set $\left(-{\Delta }_0\times {10}^{-3}\le E\le {\Delta }_0\times {10}^{-3}\right)$, and $N_{normal}=60$ (the number of pints in normal region). As Fig.A1 shows, the total current ($I_L-I_R$) behaves as, $I_0{\mathrm{sin} (\Delta \varphi )\ }$ when the phase $\varphi $ ($-\varphi $) is applied to the left (right) lead at the range $\left(0,\pi \right)$. The behavior is attributed to the Andreev bound states and shows that he transmission value $\tau <1$ (i.e., anti-crossing regime) [48,56]. It should be noted that, in addition to the above mentioned constraint related to the lattice parameter, the length of the normal region (i.e., the number of lattice points) is also important to satisfy the condition $L\gg \xi $ ($L\ll \xi $) for long (short) SC/N/SC junction where $\xi =\frac{\hslash v_F}{\pi {\Delta }_0}$ is the superconducting coherence length, with $v_F$ is the Fermi velocity [57].

\begin{figure}
\centering{}
\includegraphics[width=.8\linewidth]{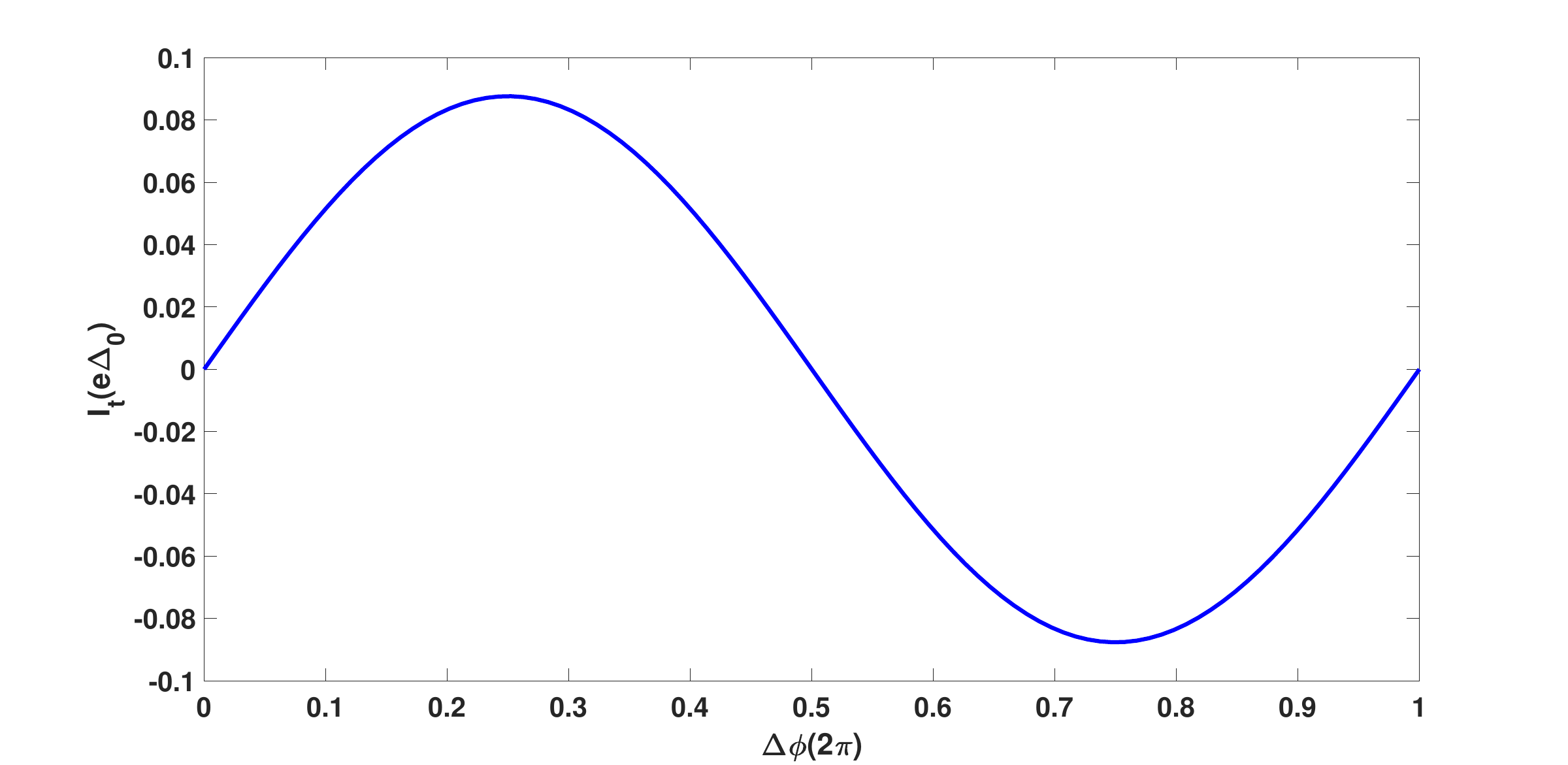}
\caption{\textbf{(Color online) the total current ( ${\boldsymbol{I}}_{\boldsymbol{L}}\boldsymbol{-}{\boldsymbol{I}}_{\boldsymbol{R}}$\textbf{) versus the applied phase to leads in superconductor-normal-superconductor structure. }${\boldsymbol{\Delta }}_{\boldsymbol{L}\boldsymbol{(}\boldsymbol{R}\boldsymbol{)}}\boldsymbol{=}{\boldsymbol{\Delta }}_{\boldsymbol{0}}{\boldsymbol{e}}^{\boldsymbol{i}\boldsymbol{\varphi }\boldsymbol{(-}\boldsymbol{\varphi }\boldsymbol{)}}$\textbf{, }$\boldsymbol{\Delta }\boldsymbol{\varphi }\boldsymbol{=}\boldsymbol{2}\boldsymbol{\varphi },$\textbf{ (}$\boldsymbol{\varphi }\boldsymbol{\in }\boldsymbol{(}\boldsymbol{0},\boldsymbol{\pi }\boldsymbol{)}$\textbf{), and }${\boldsymbol{N}}_{\boldsymbol{normal}}\boldsymbol{=}\boldsymbol{60}\boldsymbol{\ }\boldsymbol{\mathrm{(}}\boldsymbol{\mathrm{the}}\boldsymbol{\mathrm{\ }}\boldsymbol{\mathrm{number}}\boldsymbol{\mathrm{\ }}\boldsymbol{\mathrm{of}}\boldsymbol{\mathrm{\ }}\boldsymbol{\mathrm{pints}}\boldsymbol{\mathrm{\ }}\boldsymbol{\mathrm{in}}\boldsymbol{\mathrm{\ }}\boldsymbol{\mathrm{normal}}\boldsymbol{\mathrm{\ }}\boldsymbol{\mathrm{region}}\boldsymbol{\mathrm{).}}$\textbf{}}}
\textbf{\label{fig:{FigA1}}}
\end{figure}  

\begin{figure}
\centering{}
\includegraphics[width=.8\linewidth]{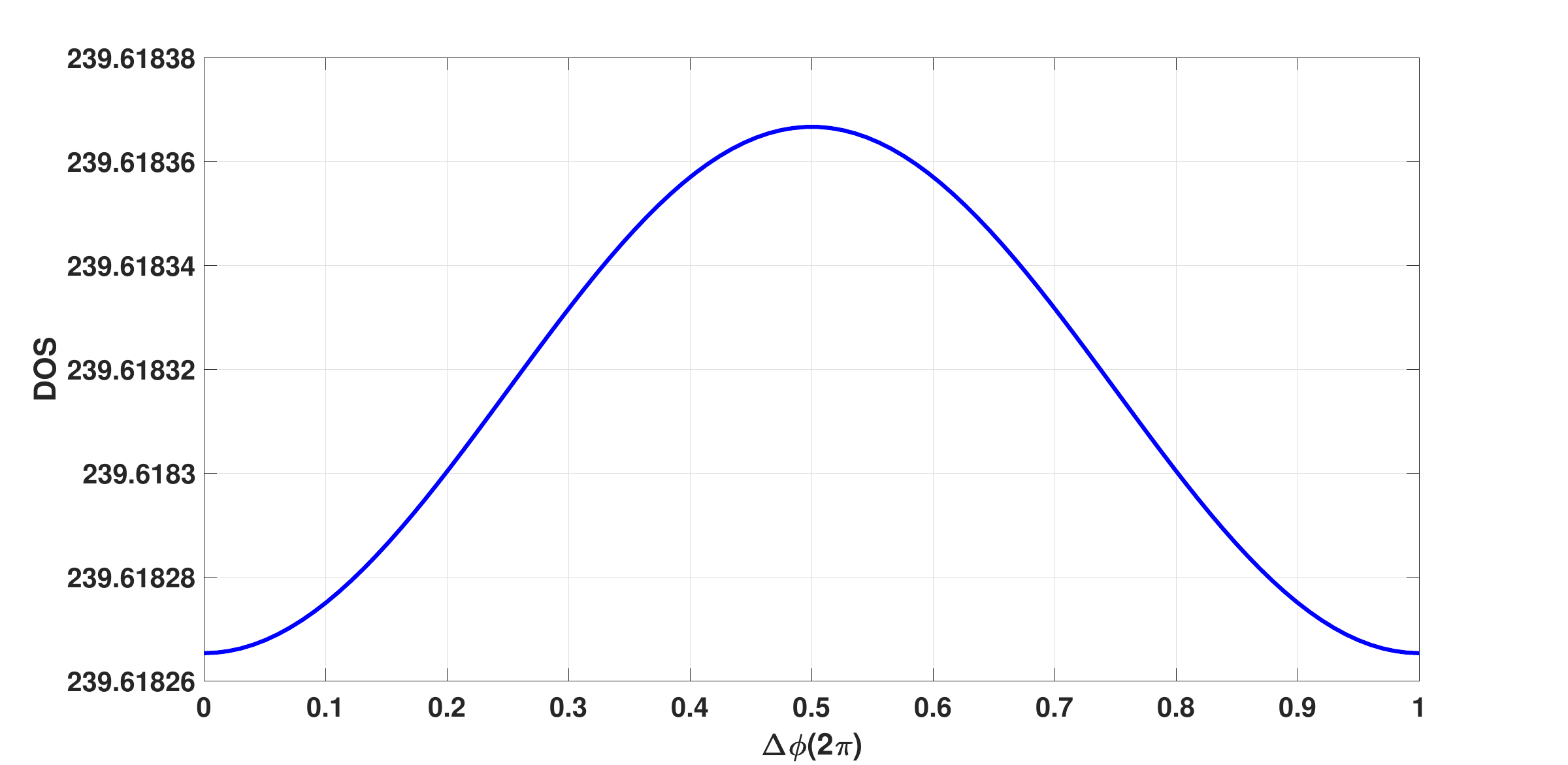}
\caption{\textbf{(Color online) The total density of state versus carrier (electrons plus holes) energy when the applied phase to the left (right) lead,  $\boldsymbol{\varphi }\boldsymbol{(-}\boldsymbol{\varphi }\boldsymbol{)}$\textbf{ changes at the range (}$\boldsymbol{0}\boldsymbol{-}\boldsymbol{\pi }$\textbf{). This behavior of DOS shows that the Andreev bound states is formed when }$\left(\boldsymbol{-}{\boldsymbol{\Delta }}_{\boldsymbol{0}}\boldsymbol{\times }{\boldsymbol{10}}^{\boldsymbol{-}\boldsymbol{3}}\boldsymbol{\le }\boldsymbol{E}\boldsymbol{\le }{\boldsymbol{\Delta }}_{\boldsymbol{0}}\boldsymbol{\times }{\boldsymbol{10}}^{\boldsymbol{-}\boldsymbol{3}}\right)$\textbf{ and the condition }$\boldsymbol{t}\boldsymbol{\gg }\boldsymbol{\mu }\boldsymbol{\gg }{\boldsymbol{\Delta }}_{\boldsymbol{0}}$\textbf{ is satisfied.}}}
\textbf{\label{fig:{FigA2}}}
\end{figure}  

\newpage

\noindent \textbf{Data availability}

\noindent  The data that supports the findings of this study are available inside its text.~

\textbf{References:}

\noindent [1] Thomas S. Kuhn, ``The Structure of Scientific Revolutions'' (Chicago: University of Chicago Press, 1970, 2${}^{nd}$ ed.).

\noindent [2] Iulia Georgescu, ``The DiVincenzo criteria 20 years on'', Nature Reviews Physics \textbf{2}, 666 (2020).

\noindent      [3] R. P. Feynman, ``Simulating Physics with Computers'', Int. Theor. Phys.    

\noindent      \textbf{21}, 467 (1982).

\noindent [4] W. Asavanant and A. Furusawa, ``Optical Quantum Computers'' (AIP Publishing, 2022).

\noindent [5] Juan Jose Gercia Ripoll, ``Quantum Information and Quantum Optics with Superconducting Circuits'' (Cambridge University Press, 2022).

\noindent [6] Tudor D. Stanescu, ``Introduction to Topological Quantum Matter and Quantum Computation'' (CRC Press, 2016).

\noindent [7] A. Furusawa, J. L. S{\o}rensen, S. L. Braunstein, C. A. Fuchs, H. J. Kimble, and E. S. Polzik, ``Unconditional Quantum Teleportation'', Science \textbf{282}, 706 (1998).

\noindent [8] M. Yukawa, R. Ukaei, P. van Look, and A. Furusawa, ``Experimental generation of four-mode continuous cluster states'', Phys. Rev. A \textbf{78}, 012301 (2008).

\noindent [9] S. Yokoyama, R. Ukai, S. C. Armstrong, C. Sornphiphatphong, T. Kaji, S. Suzuki, Jun-ichi Yoshikawa, H. Yonezawa, N. C. Menicucci, and  A. Furusawa, ,''Ultra-large-scale continuous  -variable cluster states multiplexed in time domain'', Nature Photonic \textbf{7}, 982 (2013).

\noindent [10] R. N. Alexander, S. Yokoyama, A. Furusawa, and N. C. Menicucci, ``Universal quantum computation with temporal-mode bilayer square lattice'', Phys. Rev. A \textbf{97}, 032302 (2018).

\noindent [11] S. Takeda and A. Furusawa, ``Universal quantum computing with measurement-induced continuous-variable gate sequence in a loop-based architecture'', Phys. Rev. Lett. \textbf{119}, 120504 (2017).

\noindent [12] S. Tekeda, K. Takase, and A. Furusawa, ``On-demand photonic entanglement synthesizer'', Science Adv. \textbf{5}, eaaw4530 (2019).

\noindent [13] H. P. Breuer and F. Petruccione, ``The Theory of Open Quantum Systems'' (Oxford University Press, 2002).

\noindent [14] M. Schlosshauer, ``Decoherence and the quantum to classical transition'' (Springer, 2007).

\noindent [15] S. Cong, ``Control of Quantum Systems: Theory and Methods'' (John Wiley \& Sons, 2014).

\noindent  [16] E. Cornfeld and S. Carmeli, ``Tenfold topology of crystals: Unified classification of crystalline topological insulators and superconductors'', Phys. Rev. Research \textbf{3}, 013052 (2021).

\noindent [17] Z. Xiao, R. Shindou, and K. Kawabata, ``Universal hard-edge statistics of non-Hermitian random matrices'', Phys. Rev. Research \textbf{6}, 023303 (2024).

\noindent [18] A. Yu Kitaev, ``Fault-tolerant quantum computation by anyons'', Annals of Physics \textbf{303}, 2 (2003).

\noindent [19] A. Yu Kiaev and C. Laumann, ``Topological Phases and Quantum Computation'' arXiv: 0904.2771v1(2009).

\noindent       [20] Alberto Lerda, ``Anyons: Introduction to Fractional Statistics in Two Dimensions'' (Springer, 1992). 

[21] Sumathi Rao, ``Introduction to abelian and non-abelian anyons'', arXiv:1610.09260 (2016).

[22] M. T. Deng, S. Vaitiek\.{e}nas, E. B. Hansen, J. Danon, M. Leijnse, K. Flensberg, J. Nyg{\aa}rd, P. Krogstrup, and C. M. Marcus1, ``Majorana bound state in a coupled quantum-dot hybrid-nanowire system'', Science \textbf{354}, 6319 (2016).

\noindent      [23] Stevan Nadj-Perge, Ilya K. Drozdov, Jian Li, Hua Chen, Sangjun Jeon, Jungpil Seo, Allan H. MacDonald, B. Andrei Bernevig, and Ali Yazdani, ``Observation of Majorana fermions in ferromagnetic atomic chains on a superconductor'', Sciencexpress 10.1126/science.1259327 (2014).

\noindent       [24] Qing Lin He, Lei Pan, Alexander L. Stern, Edward C. Burks, Xiaoyu Che, Gen Yin, Jing Wang, Biao Lian, Quan Zhou, Eun Sang Choi, Koichi Murata, Xufeng Kou, Zhijie Chen, Tianxiao Nie, Qiming Shao, Yabin Fan, Shou-Cheng Zhang, Kai Liu, Jing Xia, and Kang L. Wang, ``Chiral Majorana fermion modes in a quantum anomalous Hall insulator--superconductor structure'' Science \textbf{357}, 294--299 (2017).

\noindent        [25] Chun-Xiao Liu, Jay D. Sau, Tudor D. Stanescu, and S. D. Sarma, ``Andreev bound states versus Majorana bound states in quantum dot-nanowire-superconductor hybrid structures: Trivial versus topological zero-bias conductance peaks'', Phys. Rev. B \textbf{96}, 075161 (2017).

\noindent 

\noindent        [26] H. J. Suominen, M. Kjaergaard, A. R. Hamilton, J. Shabani, C. J. Palmstr{\o}m, C. M. Marcus, and F. Nichele, ``Zero-Energy Modes from Coalescing Andreev States in a Two-Dimensional Semiconductor-Superconductor Hybrid Platform'' Phys. Rev. Lett. \textbf{119}, 176805 (2017).

\noindent 

\noindent         [27] D. Wang, L. Kong, P. Fan, H. Chen, S. Zhu, W. Liu, L. Cao, Y. Sun, S. Du, J. Schneeloch, R. Zhong, G. Gu, L. Fu, H. Ding, and H. J.Gao, ``Evidence for Majorana bound states in an iron-based superconductor'' Science \textbf{362}, 333  (2018).~

\noindent        [28] H. Zhang, C. X. Liu, S. Gazibegovic, D. Xu, J. A. Logan, G. Wang, N. van Loo, J. D. S. Bommer, M. W. A. de Moor, D. Car, R. L. M. Op het Veld, P. J. van Veldhoven, S. Koelling, M. A. Verheijen, M. Pendharkar, D. J. Pennachio, B. Shojaei, J. S. Lee, C. J. Palmstr{\o}m, E. P. A. M. Bakkers, S. D. Sarma, and L. P. Kouwenhoven, ``Quantized Majorana conductance'' Nature \textbf{0}, 1 (2018).

\noindent 

\noindent        [29] R. Singh and B. Muralidharan, ``Conductance Spectroscopy of Majorana Zero Modes in Superconductor-Magnetic Insulator Nanowire Hybrid Systems'' Communications Physics~\textbf{6}, 36~(2023).

\noindent        [30] M. Kayyalha, D. Xiao1, R. Zhang, J. Shin, J. Jiang, F. Wang, Y. F. Zhao, R. Xiao, L. Zhang, K. M. Fijalkowski, P. Mandal, M. Winnerlein, Ch. Gould, Q. Li, L. W. Molenkamp, M. H.W. Chan, N. Samarth, and C. Z. Chang, ``Absence of evidence for chiral Majorana modes in quantum anomalous Hall-superconductor devices'' Science \textbf{367}, 64 (2020).

\noindent        [31] Microsoft Azure Quantum, ``Interferometric single-shot parity measurement in InAs--Al hybrid devices'' Nature \textbf{638}, 20 (2025).

\noindent        [32] Microsoft Quantum, ``Roadmap to fault tolerant quantum computation using topological qubit arrays'' arXiv:2502.12252v2 [quant-ph] (2025).

\noindent 

\noindent        [33] D. Xiao,~G. B. Liu,~W. Feng,~X. Xu, and~W. Yao, ``Coupled Spin and Valley Physics in Monolayers of~MoS${}_{2}$~and Other Group-VI Dichalcogenides'', Phys. Rev. Lett.~\textbf{108},  196802 (2012).

\noindent [34] N. F. Q. Yuan,~K. F. Mak, and~K.\ T. Law, ``Possible Topological Superconducting Phase of MoS${}_{2}$'', Phys. Rev. Lett.~\textbf{113}, 097001 (2014).

\noindent 

\noindent [35] H. Simchi, ``A simple tight-binding approach to topological superconductivity in monolayer MoS2'', Chinese Phys. B~\textbf{29},~027401 (2020).

\noindent 

\noindent [36] H. Simchi,~M. Simchi,~M. Fardmanesh~and~F. M. Peeters, ``Phase transition and field effect topological quantum transistor made of monolayer MoS2'', J. Phys.: Condens. Matter~\textbf{30},~235303 (2018).

\noindent 

\noindent [37] M. X. Deng, G. Y. Qi, W. Luo, R. Ma, R. Q. Wang, R. Shen, L. Sheng, and D. Y. Xing, ``Superconducting states and Majorana modes in transition-metal dichalcogenides under inhomogeneous strain'', Phys. Rev. B \textbf{99}, 085106 (2019).

\noindent 

\noindent [38] Y. T. Hsu,~ A. Vaezi,~M. H. Fischer~and E. A. Kim, ``Topological superconductivity in monolayer transition metal dichalcogenides'', Nature Communications~volume~\textbf{8}, 14985~(2017).

\noindent 

\noindent 

\noindent [39] M. S. Sahoo, and~M. Franz, ``High-Temperature Majorana Zero Modes'', Phys. Rev. Lett.~\textbf{128}, 137002 (2022).

\noindent 

\noindent [40] R. Zhang, I-Ling Tsai, J. Chapman, E. Khestanova, J. Waters, and I. V. Grigorieva, ``Superconductivity in Potassium-Doped Metallic Polymorphs of MoS${}_{2}$'', Nano Lett. \textbf{16}, 629 (2016).

\noindent 

\noindent [41] R. D. Septianto, A. P. Romagosa, Y. Dong, H. Matsuoka, T. Ideue, Y. Majima, and Y. Iwasa, ``Gate-Controlled Potassium Intercalation and Superconductivity in Molybdenum Disulfide'' Nano Lett. \textbf{24}, 13790 (2024).

\noindent 

\noindent 

\noindent [42] Y. Saito, Y. Nakamura, M. S. Bahramy, Y. Kohama, J. Ye, Y. Kasahara, Y. Nakagawa, M. Onga, M. Tokunaga, T. Nojima, Y. Yanase, and Y. Iwasa, ``Superconductivity protected by spin--valley locking in ion-gated MoS${}_{2}$'', Nature Phys. \textbf{7}, 3580 (2015).

\noindent 

\noindent 

\noindent [43] B. Uchoa and Y. Barlas, ``Superconducting States in Pseudo-Landau-Levels of Strained Graphene'', Phys. Rev. Lett. \textbf{111}, 046604 (2013).

\noindent 

\noindent [44] L. Majidi,~H. Rostami, and~R. Asgari, ``Andreev reflection in monolayer MoS${}_{2}$'', Phys. Rev. B~\textbf{89}, 045413 (2014).

\noindent 

\noindent [45] H.~Goudarzi,~M.~Khezerlou,~and S.F.~Ebadzadeh, ``Andreev reflection and subgap conductance in monolayer MoS2~ferromagnet/s and d-wave superconductor junction'', Superlattices and Microstructures \textbf{93}, 73 (2016).

\noindent 

\noindent [46] C. Bai1,~Y. Zou,~W. K. Lou, and~K. Chang, ``Pure valley- and spin-entangled states in a~MoS${}_{2}$-based bipolar transistor'', Phys. Rev. B~\textbf{90}, 195445 (2014).

\noindent 

\noindent [47] J.O. Island, ``Quantum transport in superconducting hybrids'' (PhD Thesis, Delft University of Technology, 2016).

\noindent [48] Z. Y. Zeng, B. Li, and F. Claro, ``Electronic transport in hybrid mesoscopic structures: A nonequilibrium Green function approach'', Phys. Rev. B 8, 115319 (2003).

\noindent 

\noindent [49] P. Sriram, S. S. Kalantre, K. Gharavi, J. Baugh, and B. Muralidharan, ``Supercurrent Interference in Semiconductor Nanowire Josephson Junctions'', Phys. Rev. B \textbf{100}, 155431 (2019).

\noindent 

\noindent [50] Y. Xu, S. Uddin, J. Wang, Z. Ma, and J. F. Liu, ``Electrically modulated SQUID with a single Josephson junction coupled by a time reversal breaking Weyl semimetal thin film'', Phys. Rev. B \textbf{97}, 035427 (2018).

\noindent 

\noindent [51] C. Duse,~P. Sriram,~K. Gharavi,~J. Baugh,~B. Muralidharan, ``Role of dephasing on the conductance signatures of Majorana zero modes'' J. Phys. Condens. Matt. 33, 365301 (2021).

\noindent 

\noindent [52] A. Kejriwal and B. Muralidharan, ``Can non-local conductance spectra conclusively signal Majorana zero modes? Insights from von Neumann entropy'', arXiv:2112.02235v1(2021).

\noindent 

\noindent [53] IEEE, ``International Roadmap of Devices and Systems (IRDS${}^{TM}$)'' (IEEE, 2024 edition).

\noindent 

\noindent [54] Jian-Xin Zhu, ``Bogoliubov-de Gennes Methods and Its Applications'' (Springer, 2016).

\noindent 

\noindent [55] Y. M. Xiao, W. Xu, B. Van Duppen, and F. M. Peeters, ``Infrared to terahertz optical conductivity of n-type and p-type monolayer MoS${}_{2}$ in the presence of Rashba spin-orbit coupling'' Phys. Rev. B \textbf{94}, 155432 (2016).

\noindent 

\noindent [56] M. P. Lopez Sancho, J. M. Lopez Sancho, and J. Rubio, ``Highly convergent schemes for the calculation of bulk and surface Green functions'' I. Phys. F: Met. Phys. \textbf{15} 851 ((1985).

\noindent 

\noindent [57] Jorge Cayao, ``Hybrid superconductor-semiconductor nanowire junctions as useful platforms to study Majorana bound states'' (PhD Thesis, Autonomous University of Madrid, 2016) arXiv:1703.07630v1 (2017). 

\noindent [58] Kh. Shakouri, H. Simchi, M. Esmaeilzadeh, H. Mazidabadi, and F. M. Peters, ``Tunable spin and charge transport in silicone nanoribbon'', Phys. Rev. B \textbf{92}, 035413 (2015).

\noindent [59] Supriyo Datta, ``Quantum transport: Atom to Transistor'' (Cambridge University Press, 2005).

\noindent

\noindent

\end{document}